\documentclass[12pt]{article}
\usepackage{graphics, color}
\usepackage{graphicx}
\usepackage{amssymb}
\usepackage{youngtab}
\usepackage{hyperref} 
\usepackage{psfrag}

\newcommand{\sect}[1]{\section{#1}\setcounter{equation}{0}}

\def\gsim{\, \rlap{$>$}{\lower 1.1ex\hbox{$\sim$}}\,}
\def\lsim{\, \rlap{$<$}{\lower 1.1ex\hbox{$\sim$}}\,}

\newcommand{\vp}{{\vphantom\dagger}}

\newcommand{\ra}{\rightarrow}

\newcommand{\be}{\begin{equation}}
\newcommand{\ee}{\end{equation}}
\newcommand{\ba}{\begin{eqnarray}}
\newcommand{\ea}{\end{eqnarray}}
\newcommand{\bi}{\begin{itemize}}  
\newcommand{\ei}{\end{itemize}}

\newcommand{\Tr}{{\rm Tr}}
\newcommand{\nn}{\nonumber}
\newcommand{\mo}{{-1}} 
\newcommand{\oo}{\frac{1}}
\newcommand{\f}{\frac}
\newcommand{\half}{\frac{1}{2}}
\renewcommand{\bar}{\overline}

\begin{document}


\begin{titlepage}
\rightline{\tt NSF-KITP-08-115}
\bigskip\bigskip\bigskip\bigskip\bigskip
\centerline{\Large Matrix Models for the Black Hole Information Paradox}
\bigskip\bigskip
\centerline{{\bf Norihiro Iizuka}\footnote{\tt iizuka@kitp.ucsb.edu}}
\medskip
\centerline{{\bf Takuya Okuda}\footnote{\tt takuya@kitp.ucsb.edu}}
\medskip
 \centerline{{\bf Joseph Polchinski}\footnote{\tt joep@kitp.ucsb.edu}}
\bigskip
\centerline{\em Kavli Institute for Theoretical Physics}
\centerline{\em University of California}
\centerline{\em Santa Barbara, CA 93106-4030, USA}
\bigskip

\bigskip


\begin{abstract}
We study various matrix models with a charge-charge interaction as toy models of the  gauge dual of the AdS black hole.  These models show a continuous spectrum and power-law decay of correlators at late time and infinite $N$, implying information loss in this limit.   At finite $N$, the spectrum is discrete and correlators have recurrences, so there is no information loss.  We study these models by a variety of techniques, such as Feynman graph expansion, loop equations, and sum over Young tableaux, and we obtain explicitly the leading $1/N^2$ corrections for the spectrum and correlators.  
These techniques are suggestive of possible dual bulk descriptions. 
At fixed order in $1/N^2$
the spectrum remains continuous and
no recurrence occurs, so information loss persists.
However, the interchange of the long-time and large-$N$ limits is subtle and requires further study.
\end{abstract}
\end{titlepage}
\baselineskip = 16pt

\setcounter{footnote}{0}


\sect{Introduction}

The black hole information paradox~\cite{Hawking:1976ra} has been a fruitful thought experiment, leading in particular to the discovery of gauge/gravity duality~\cite{Maldacena:1997re}.  This duality in turn
provides a nonperturbative definition of string theory as quantum gravity in AdS backgrounds
in terms of unitary quantum mechanics,
and  implies that the information escapes with the Hawking radiation, but important questions remain.  In particular, how does the argument for information loss, based on the low energy effective field theory of gravity, break down?

Ref.~\cite{Iizuka:2008hg}, building on ideas of Refs.~\cite{Iizuka:2001cw,Festuccia:2006sa}, presented a simple model in which this might be studied further.  
 The quantum mechanical system of a single large-$N$ matrix oscillator and a single fundamental oscillator displays the key property~\cite{Maldacena:2001kr} that information is lost at infinite $N$ but not for $N$ finite.  
Since $1/N^2$ is proportional to $G_N$ in gravity,
this result demonstrates that quantum gravity effects  
are crucial to avoid information loss.
In the planar limit the Schwinger-Dyson equation for this model closes, and moreover can be reduced to a recursion relation with respect to frequency~\cite{Iizuka:2008hg}; analytic arguments, and numerical solution, then confirm the desired properties.  However, a complete analytic solution seems difficult even in the planar limit, and a systematic study of the $1/N^2$ corrections is even more difficult.

In the present paper we present more tractable models that have the same degrees of freedom, but where the previous trilinear interaction~\cite{Iizuka:2008hg} is replaced by one that is quartic in the oscillators but quadratic in the $U(N)$ charges.  In Sec.~2 we motivate this by showing that the weak-coupling limit of the trilinear model leads to an effective charge-charge interaction.  We then consider more general charge-charge interactions, and present the solution in the planar limit.  Like the trilinear model, this displays the essential feature of a continuous spectrum at infinite $N$.  
The Schwinger-Dyson equation is algebraic, and the decay of the planar two-point function is power-law rather than exponential at late times. Still, information is lost at large $N$, allowing us to address the paradox in a simpler setting.

Our ultimate goal is to see whether the preservation of information might be reflected in the large-time and/or large order (in $1/N^2$) behaviors of the perturbative expansion.  We are also interested in developing the analog of a bulk string/gravity description, to see how information is preserved in this language.  To these ends we attempt to solve the model in several different frameworks.

In Sec.~3 we further develop the graphical approach, beyond the planar limit.   We find explicitly the first non-planar amplitude, summing all Feynman graphs that have the topology of a disk with a handle.
Our explicit computation of the first $1/N^2$ correction
demonstrates that the spectrum remains continuous.
No recurrence occurs and information loss persists.

In Sec.~4 we analyze the theory in terms of loop equations, recursive relations for expectation values of operators.  This allows us to recover the planar term and first non-planar correction, and generalizes efficiently to higher orders.

In Sec.~5 we show that the correlator for this theory can be written for any $N$ in terms of a sum over Young tableaux.  The planar limit is recovered as a large-$N$ saddle point.

In Sec.~6 we study various models in which the matrix oscillator is generalized to a rectangular $N \times K$ matrix.  This allows for several limits of large and small $N$ and $K$ where a more explicit solution is possible.

In Sec.~7 we discuss the results and future directions.  In particular we examine the possibility that one might see signs of the long-time recurrences in the behavior of the $1/N^2$ expansion.

\sect{Charge-charge models}

\subsection{Models}
\label{models}

First we recall the trilinear model~\cite{Iizuka:2008hg}.  In terms of lowering operators $A_{ij}$ and $ a_i$ for the adjoint and fundamental, the Hamiltonian is
\begin{equation}
H = m \,{\rm Tr}(A^\dagger A) + M a^\dagger a  +{g} a^\dagger X a
 + c\, a^\dagger a (a^\dagger a  - 1)
\ . \label{ham}
\end{equation}
where $X = (A+A^\dagger)/\sqrt{2m}$.
The final term is needed to stabilize the system, but has been arranged to vanish in the relevant sectors $a^\dagger a = 0,1$.   We take $M$ to be large so that $a^\dagger a $ is essentially zero in the thermal ensemble.

\begin{figure}
\center \includegraphics[width=13cm]{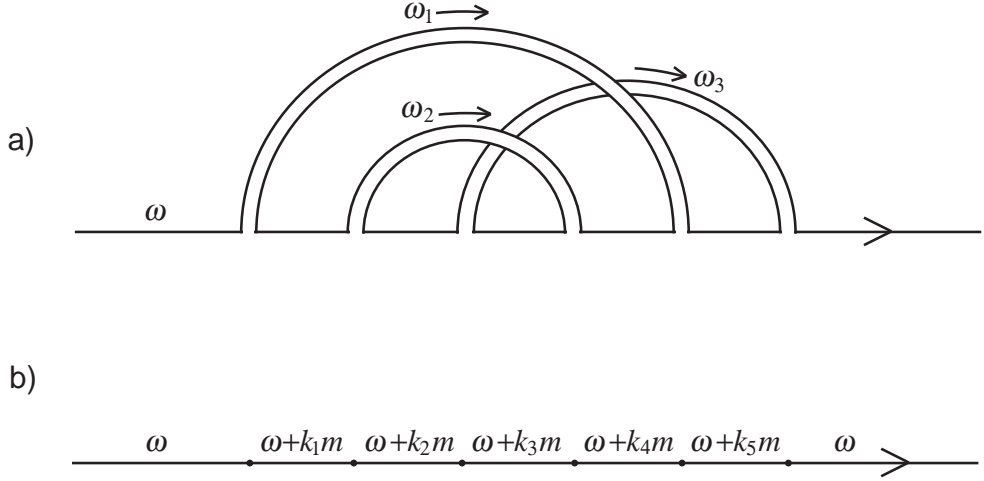}   
\caption[]{a) Typical Feynman graph for the fundamental correlator~(\ref{correlator}) in the trilinear model (\ref{ham}).  The only poles in the lower half $\omega_{1,2,3}$-planes are from the adjoint propagators. After evaluating residues this leaves b), where $k_{r+1} = k_r \pm 1$.  The most singular graphs alternate between $k = 0$ and $k = \pm 1$.} 
\label{graph}
\end{figure}

A typical Feynman graph for 
\begin{equation}
e^{iM(t-t')} \left\langle {\rm T}\, a\vp_i(t) a_j^\dagger(t') \right\rangle_T
\equiv \delta_{ij} G(T,t-t')
 \label{correlator}
\end{equation}
is shown in Fig. \ref{graph}. 
The free thermal propagators are
\begin{eqnarray}
\tilde G_0(T,\omega) &=& \tilde G_0(\omega) = \frac{i}{\omega + i\epsilon}\ , \nonumber\\
 \tilde K_0(T,\omega) &=& \frac{i}{1 -y} \left( \frac{1}{\omega^2 - m^2 + i\epsilon}
- \frac{y}{\omega^2 - m^2 - i\epsilon} \right) \ ,\quad y = e^{-{m/T}}\ . \label{ktherm}
\end{eqnarray}
The only singularities in the lower-half $\omega_{1,2,3}$ planes are from the adjoint propagators, at $\pm m - i\epsilon$, so we can evaluate these integrals by residues, leaving the fundamental propagators at momenta $\omega + k m$ for various integers $k$.  Thus, an individual Feynman graph gives only poles on the real axis, as in the discussion in Ref.~\cite{Festuccia:2006sa}. However, it is shown in Ref.~\cite{Iizuka:2008hg} that this cannot be a property of the full planar propagator for any nonzero $g$.  The point is that perturbation theory is singular, because higher orders of perturbation theory give higher order poles.  The most singular graphs as $\omega$ goes to its on-shell value 0 are those in which $k$ alternates between $\pm 1$ and 0, since this gives the maximum number of poles.  

Thus we can capture the most singular graphs by integrating out the propagators at $k = \pm 1$, leaving an effective quartic interaction with two adjoints and two fundamentals.\footnote{
This is equivalent to dropping smaller terms
in the recursion relations for $\tilde G$, as was done in~\cite{Iizuka:2008hg}.
}  In other words, we must do degenerate perturbation theory, because states with the same total number of $A^\dagger$ excitations are degenerate in the free theory.  Thus
\begin{eqnarray}
H_{\rm eff} &=& H_{\rm int} \frac{1}{E_0 - H_0}  H_{\rm int} 
\nonumber\\
&=&
\frac{g^2}{2m^2} (a_i^\dagger
A_{ij}^\dagger a^\vp_j\, a_k^\dagger A^\vp_{kl} a^\vp_l - a_i^\dagger A^\vp_{ij} a^\vp_j\, a_k^\dagger A_{kl}^\dagger a^\vp_l)  + O(AA, A^\dagger A^\dagger) \nonumber\\
&=& 
\frac{g^2}{2m^2} (a_i^\dagger
A_{ij}^\dagger  A^\vp_{jl} a^\vp_l - a_i^\dagger A^\vp_{ij}  A_{jl}^\dagger a^\vp_l)  + O(AA, A^\dagger A^\dagger) \ .
\end{eqnarray}
In the last line we have projected down to the $a^\dagger a = 1$ sector (which is annhilated by $a_j a_l$), as relevant for $G$.

The final interaction is simply a coupling of the $U(N)$ charges of the fundamental and the adjoint,
\begin{equation}
H_{\rm int}=-\frac{g^2}{2m} q_{li} {\cal Q}_{il}  \label{qQ}
\end{equation}
with
\begin{equation}
 q^\vp_{li} = - a_i^\dagger a^\vp_l\ ,\quad {\cal Q}^\vp_{il} = A_{ij}^\dagger  A^\vp_{jl} - A^\vp_{ij}  A_{jl}^\dagger \ .
\end{equation}
Thus, the energy can be expressed in terms of a difference of quadratic Casimirs.  This implies a large degeneracy; nevertheless, the model will still have enough mixing to produce a continuous spectrum in the large-$N$ limit.

We could generalize by giving independent coefficients to the two terms in $Q_{il}$.  In fact, these two terms separately generate commuting $U(N)$'s, the first acting on the left index of $A^\dagger_{ij}$ and the second on the right index.  We obtain a slightly simpler model (in terms of its Schwinger-Dyson equation) by keeping only one term,
\begin{equation}
\label{ourHint}
H_{\rm int}=
-h q^\vp_{li} Q_{il}\ ,\quad Q_{il} = A_{ij}^\dagger  A^\vp_{jl}\ .  \label{tilde}
\end{equation}

\subsection{Planar solution}

\begin{figure}
\center \includegraphics[width=14cm]{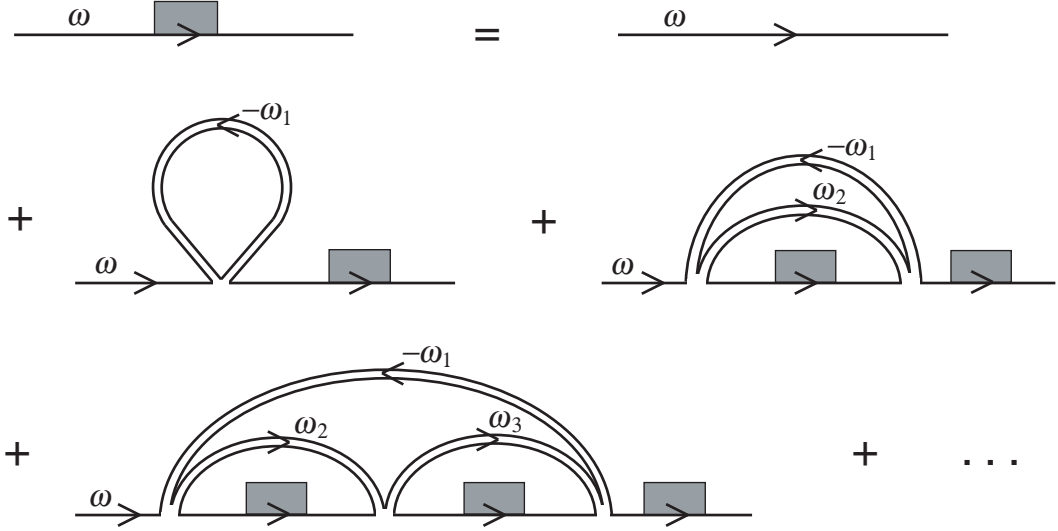}   
\caption[]{Planar Feynman graphs for the correlator~(\ref{correlator}) in the model~(\ref{tilde}).  The shaded rectangles mark the full planar propagators.  Arrows point from creation operators toward annihilation operators.  The graphs for $n = 0,1,2$ are shown.} 
\label{planarSD-fig}
\end{figure}

Now let us solve these in the planar approximation.  For the $qQ$ model~(\ref{tilde}) the nontrivial planar graphs involve a cycle with $n+1$ vertices, giving the Schwinger-Dyson equation shown in Fig. \ref{planarSD-fig}.  
Thus,
\begin{eqnarray}
\label{planarSD}
\tilde G(T,\omega) &=& \tilde G_0(\omega) + \tilde G_0(\omega) \tilde G(T,\omega)
\sum_{n=0}^\infty S_n(T,\omega) \nonumber\\
S_n(T,\omega) &=& (-i hN)^{n+1} \int \frac{d^{n+1} \vec\omega}{(2\pi)^{n+1}} \tilde L_0(T,-\omega_{1})
\prod_{l=1}^n  \left[ \tilde G(T,\omega - \omega_{l+1} - \omega_{1}) \tilde L_0(T,\omega_{l+1}) \right] \ ,\ \ 
\end{eqnarray}
where $\tilde L_0(T,\omega)$ is the thermal $AA^\dagger$ propagator
\begin{equation}
\tilde L_0(T,\omega) = \frac{i}{1 -y} \left( \frac{1}{\omega - m + i\epsilon}
- \frac{y}{\omega - m - i\epsilon} \right) \ .
\end{equation}
At large $M$, $\tilde G(T,\omega)$ has singularities only in the lower half $\omega$-plane, and $\tilde G_0$ and $\tilde L_0$ both fall as $1/\omega$ at large frequency, so we can close the $\omega_i$ integrals in the lower half-plane and pick up residues only from $\tilde L_0$.  This gives
\begin{equation}
\label{planarfn}
S_n(T,\omega) =  y\left(\frac{-i h N}{1-y} \right)^{n+1} \tilde G(T,\omega)^n \ .
\end{equation}
The Schwinger-Dyson equation becomes
\begin{equation}
-i \omega \tilde G(T,\omega)= 1 - \frac{i y \lambda  \tilde G(T,\omega)}{1 - y + i \lambda  \tilde G(T,\omega)}\ ,
\end{equation}
or
\begin{equation}
-\omega \lambda \tilde G^2(T,\omega) + i (1-y)(\omega + \lambda) \tilde G(T,\omega) + (1-y)=0 \ , \label{SDquad}
\end{equation}
where $\lambda = h N$ is the 't Hooft coupling.  The solution is 
\begin{equation}
\label{Gzerosol}
\tilde G(T,\omega)= \frac{i(1-y)}{2\omega\lambda}\left( \lambda + \omega - \sqrt{(\omega-\omega_+)
(\omega-\omega_-)} \right)\ ,\quad
\omega_{\pm} = \lambda \frac{ 1+ y \pm 2 \sqrt{y} }{1-y}\ . \label{primed-G-SD}
\end{equation}
This has a pole of spectral weight $1-y$ at $\omega = 0$, and a cut from $\omega_-$ to $\omega_+$.  
 
For the original model~(\ref{qQ}) the only modification is the inclusion of an additional copy of each cycle but with the arrows reversed, so that
\begin{equation}
S_n(T,\omega) =  [y - (-y)^n] \left(\frac{-i \lambda}{1-y} \right)^{n+1} \tilde G(T,\omega)^n
 \ .
\end{equation}
with $\lambda = hN$.
Then
\begin{equation}
-i \omega \tilde G(T,\omega)= 1 - \frac{i y \lambda \tilde G(T,\omega)}{1 - y + i \lambda \tilde G(T,\omega)} +\frac{i \lambda \tilde G(T,\omega)}{1 - y - i y\lambda \tilde G(T,\omega)}\ .
\end{equation}
This reproduces the cubic equation for $\tilde G(T,\omega)$, Eq.~(32) of~\cite{Iizuka:2008hg}, which was obtained from the weak-coupling approximation to the recursion relation of the trilinear model.  Again, there is a branch cut on a finite segment of the real axis.

\subsection{Black hole physics}

The continuous spectrum implied by the cut in these models is the signature of a horizon.  As noted in Ref.~\cite{Iizuka:2008hg}, the cut is absent at zero temperature, and also below the Hawking-Page transition (which we can simulate in this one-matrix model by imposing the singlet constraint).  The Fourier transform of the cut gives a $t^{-3/2}$ behavior at late times.  Although this falls off more slowly than the exponential for the real black hole, it is still inconsistent with the properties of a system with finite entropy, and so with the exact correlator.  Indeed, the energies in this model are all multiples of $h = \lambda/N$, so there are regular recurrences with period $2\pi N/\lambda$.  This can be seen by writing the interaction in terms of quadratic Casimirs, as we will do in Sec.~5.1.  This time is much shorter than for a fully thermalized system, for which the energy splittings are of order $e^{-N^2}$, but still presents us with a version of the information paradox.  We should note that the more general model
\begin{equation}
H_{\rm int}=
-q^\vp_{li} (h_1  A_{ij}^\dagger  A^\vp_{jl} - h_2  A_{ij}^\vp  A^\dagger_{jl})
\ . 
\end{equation}
cannot be written in terms of commuting Casimirs for generic $h_{1,2}$, and so may give a more realistic model of the black hole.


\sect{Non-planar corrections}

\subsection{Schwinger-Dyson equation}

We now consider the full Schwinger-Dyson equation, including non-planar corrections.  We focus henceforth on the $qQ$ model~(\ref{tilde}).  It is useful first to carry out all of the loop integrations, as in the previous discussion.  The number of loops is equal to the number of adjoint propagators, so we can take the adjoint propagator momenta $\omega_i$ as integration variables.  We orient these in the direction of the arrow on the fundamental propagator, as in Figs.~1, 2.  The momentum on the fundamental propagator therefore always involves $-\omega_i$, and so this propagator contributes no poles in the lower-half $\omega_i$ plane.  Thus we can close the loop integrals in this half-plane picking up the pole at $m- i \epsilon$ for each forward propagator (one whose arrow is parallel to that on the fundamental line) and at $-m- i \epsilon$ for each backward propagator (antiparallel to the fundamental line).  Further, since each vertex contains one $A$ and one $A^\dagger$, there are always equal numbers of $+m$ and $-m$ loop momenta flowing on any internal fundamental line, so this is always at $\omega_{\rm internal} = \omega_{\rm external}$.  This is simply a repetition of the point made in Sec.~2.1, that this model isolates the propagators with $k=0$.

We therefore have the Feynman rules
\begin{eqnarray}
\longrightarrow&&\quad G_0(\omega)\ , \quad \omega = \omega_{\rm external}\ , \nonumber\\
\Longrightarrow&&\quad \frac{1}{1-y}\ , \nonumber\\
\Longleftarrow &&\quad \frac{y}{1-y}\ . \label{backfor}
\end{eqnarray}
The evaluation of the amplitudes is thus reduced to counting graphs,
\begin{equation}
\tilde{G}(T,\omega) =  \tilde G_0(\omega) \sum_{\rm graphs} \left({- i \lambda \over {1 - y}} \tilde{G_0}(\omega) \right)^v y^b  \Biggl({1\over N^2} \Biggr)^g\ ,
\end{equation}
where $v$ is the number of vertices, $b$ is the number of backward propagators, and $g$ is the genus of the graph.

We now reduce the sum over graphs by summing over certain classes of subgraph.  Start with the sum of the one-particle-irreducible (1PI) fundamental self-energy graphs $S(T,\omega,h,G_0)$, in terms of which the full Schwinger-Dyson equation reads
\ba
\tilde{G}(T,\omega) =  \tilde G_0(\omega) 
+\tilde G_0(\omega) S(T,\omega, h,G_0) \tilde G(T,\omega).
\ea  
We denote explicitly its dependence on the coupling and the functional form of the propagator for the fundamentals.  
Consider self-energy corrections on internal propagators.
Because we are assuming that the fundamental oscillator mass $M$ is large, there are no loops of the fundamental field, and so there are no corrections to the adjoint propagator, only to the fundamental propagator. 
By packaging corrections on internal fundamental propagator as $G$,
we only have to sum over graphs without them.
What we get is precisely the two-particle-irreducible (2PI) fundamental self-energy $S_{\rm 2PI}$,\footnote{A 2PI graph is one that cannot be separated into two pieces
by cutting two propagators.
Such a graph clearly cannot have an internal propagator correction.
If a graph in $S$ can be separated in this way,
at least one fundamental propagator has to be cut, because all vertices lie along a single fundamental line.  
Since the number of forward minus backward propagators emerging from any subgraph is always zero, the second propagator to be cut
must be a fundamental line, isolating the propagator corrections in between.
}
so
\begin{equation}
S(T,\omega,h,G_0) = S_{\rm 2PI}(T,\omega,h,G) \,.
\end{equation}

Next, for any vertex, there is a geometric series of graphs obtained by expanding it into $n+1$ vertices, each connected to the next by a forward adjoint propagator and a fundamental propagator.  This is illustrated in the SD equation of Fig.~2, where all of the interaction terms can be obtained from the $n=0$ term in this way.  Thus, we can sum all such series into effective vertices.  
We will refer to the propagators to be summed as 
 {\it trivial forward propagators}, a term which 
we will explain in Sec. \ref{graph-correction}.
So we can restrict the sum to graphs with no trivial forward propagators (NTF) but with the geometric series incorporated into the vertices:
\begin{equation}
S(T,\omega,h,G_0) = S_{\rm I^*}(T,\omega,h',G)\ ,\quad h' = \frac{h}{1 + i \lambda \tilde G(T,\omega)/(1-y)}\ .
\end{equation}
Here $S_{\rm I^*}$ denotes the sum of the graphs that are fully irreducible (I$^*$), i.e.\ 1PI, 2PI, {\it and} NTF.  
\begin{figure}
\center \includegraphics[width=14cm]{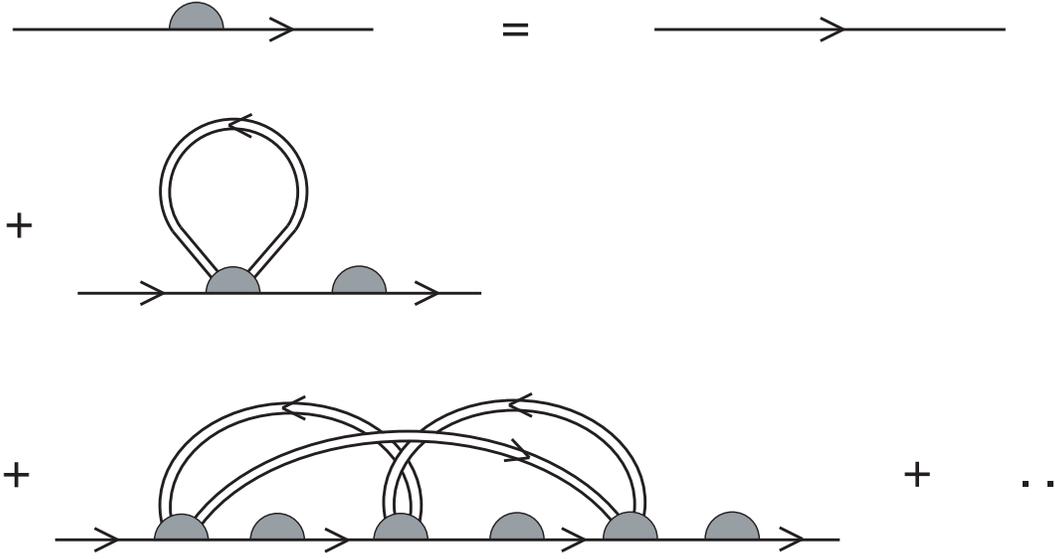}   
\caption[]{Schwinger-Dyson equation for the full propagator.  The sum runs over all I$^*$ graphs.  Lines with shaded semicircles denote full propagators.  Vertices with shaded semicircles contain the summed trivial forward propagators.  When we refer to the genus of this graph (as in Sec.~3.2, where we sum all genus one graphs) we mean the explicit genus, not taking into account the non-planar corrections implicit in the shaded circles.} 
\end{figure}

One form for the Schwinger-Dyson equation is then
\begin{equation}
\label{genericSD}
\tilde G(T,\omega) = \frac{i}{\omega} +   \frac{i}{\omega} 
S_{\rm I^*}(T,\omega, h',G)
\, \tilde G(T,\omega)\ .
\end{equation} 
The total contribution of a given I$^*$ graph to $ S_{\rm I^*}(T,\omega, h',G)\, \tilde G(T,\omega)$ is
\begin{equation}
\label{genericform}
w^v y^b N^{-2g}\ , \quad w = - i h'  N \tilde G(T,\omega)/(1-y)\ . \label{weight}
\end{equation}
The first few terms are shown in Fig.~3.  
The planar contributions have been collapsed into a single term by the summation over graphs containing trivial forward propagators, but even at the first non-planar order, the number of I$^*$ graphs is infinite.  Summing these is our next exercise.

\subsection{The \texorpdfstring{$1/N^2$}{1/N2} correction}
\label{graph-correction}

If we take the trace of the fundamental propagator, which is just 
$N\tilde G(T,\omega)$, then the planar graphs are those that can be drawn on a disk, while the $N^{-2g}$ corrections come from graphs that can be drawn on a disk with $g$ handles~\cite{'t Hooft:1974hx}.  Thus we wish to enumerate all fully reducible I$^*$ graphs that can be drawn on a disk with one handle (Fig.~4). 
\begin{figure}
\center \includegraphics[width=7cm]{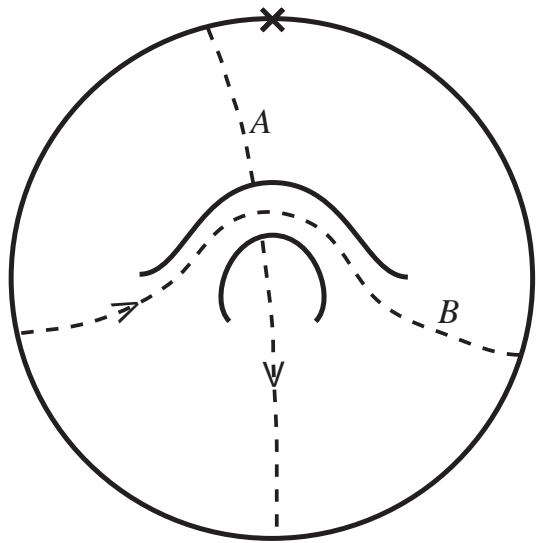}   
\caption[]{Disk with one handle.  The $\times$ indicates the point where the ends of the fundamental line are joined.  The dashed lines mark the $A$ and $B$ cycles.} 
\end{figure}
 We have marked with an $\times$ the point where the ends of the fundamental propagator are joined, because this enters into the Feynman rules.  We have also marked $A$ and $B$ cycles; every adjoint propagator is homotopic to $p A + q B$ for some integers $p$ and $q$.  For the present discussion we are not distinguishing an orientation on the adjoint propagators.

The possible 
trivial propagators, $p = q = 0$, are very limited in a 2PI graph.  
As is clear from Fig.~2, the forward propagators that have been summed into the NTF vertices
are trivial, and in fact these are the only trivial forward propagators 
as we have already mentioned.
To see this, note that a contractible forward propagator divides the Riemann surface into two pieces (that which it crosses when contracting, and that which it does not).  
It follows that if  we cut the fundamental propagators that attach to 
each end of this forward propagator, the surface is separated into two.  Because the graph is 2PI, this is only  possible if the two ends are connected by bare fundamental propagator. 
Thus it is of precisely the type summed by the NTF condition.

If a backward propagator is trivial, a similar argument shows that it also divides the Riemann surface, and so would cutting on the adjacent fundamental lines.  This is also excluded by the 2PI condition, the only exception being when the ends of the backward propagator are the first and last adjoint propagators to attach to the fundamental line, since the 1PI condition omits these adjacent propagators.  For example, all the planar graphs in Fig.~2 have such a trivial backward propagator. So for any non-planar graph, either all propagators are homotopically nontrivial, or there is a single trivial backward propagator which separates the $\times$ from the rest of the graph.

To enumerate the nontrivial propagators, we must be careful not to overcount, because the $SL(2,Z)$ modular group of the torus allows us to draw the same graph in different ways.  Therefore, we specify that as we move along the fundamental line in the direction of the arrow, the first nontrivial propagator that we encounter is homotopic to the $A$ cycle.\footnote{We can exclude a integer multiple of $A$ (which would not be modular-equivalent to $A$) because such a propagator would intersect itself.}  Similarly, we specify that the second nontrivial propagator that we meet (excluding those homotopic to the first) is homotopic to the $B$ cycle.  This fixes the modular group, and so we can count freely.  Note that there is at least one propagator along the $A$ cycle and one along the $B$ cycle, or else the graph is actually planar.

There are five possible kinds of propagator:
\begin{enumerate}
\item $n_1$ propagators which go under the handle from the
left of the marked point ($A$ cycle propagators).  As noted above, $n_1 \ge 1$. 
\item $n_2$ propagators which go directly along the handle (the $B$ cycle). 
Again, $n_2 \ge 1$ if the graph is non-planar.
\item $n_3$ propagators which go along the handle in a ``twisted'' way.  These are homotopic to $A+B$.
There are genus-one graphs without these twisted propagators, so $n_3 \ge 0$. 
\item $n_4$ propagators which go under the handle from the right of the 
marked point.  Again, these are homotopic to the $A$ cycle, but there need not be propagators of this in this group: $n_4 \ge 0$. 
\item $n_5$ trivial backward propagators as discussed above, where $n_5 \in \{0,1\}$.
\end{enumerate} 
In Fig.~5 we depict in various ways the case $n_1 = 2$, $n_2 = n_3 = n_4 = n_5 = 1$.
\begin{figure}
\center \includegraphics[width=14cm]{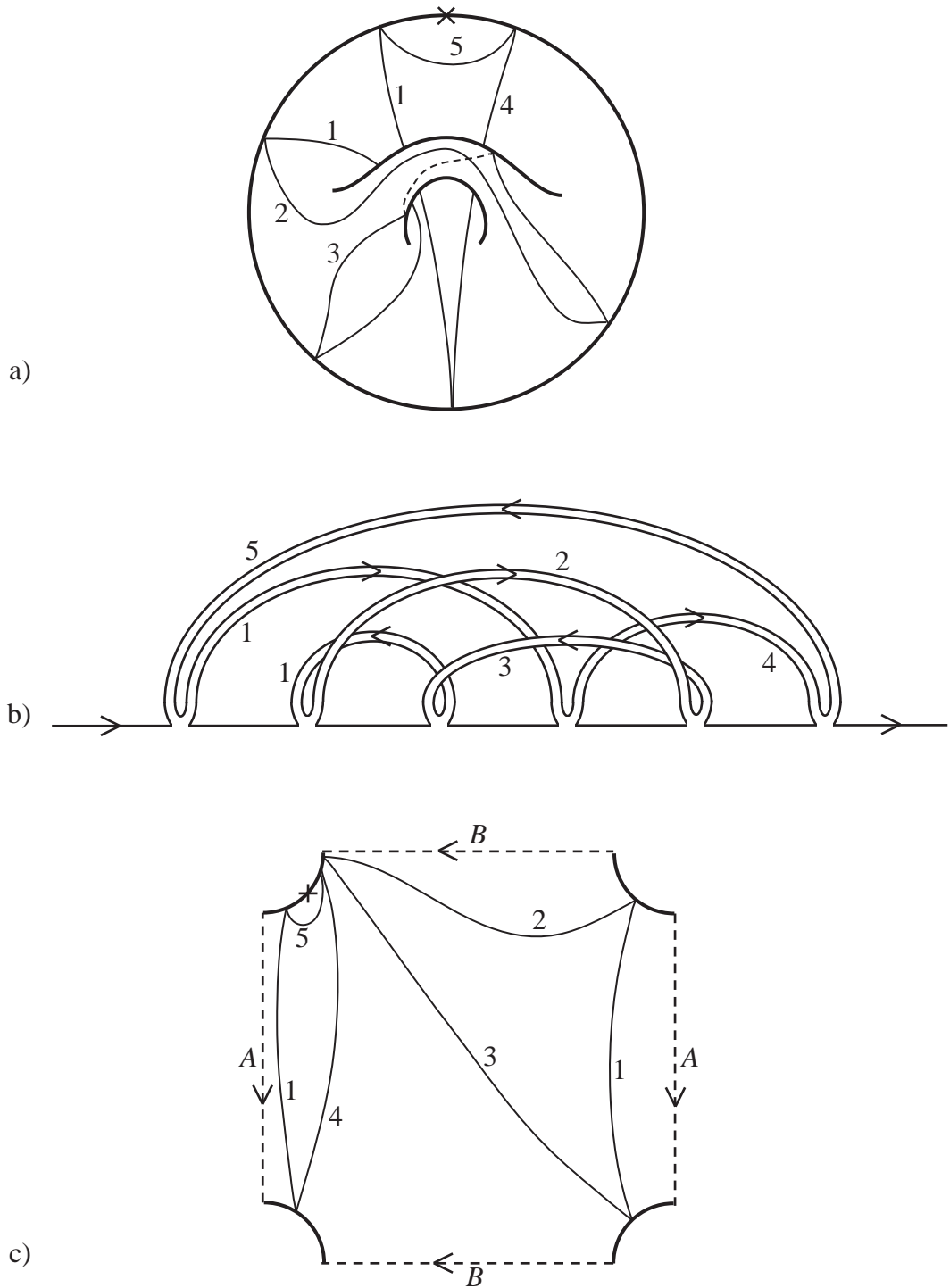}   
\caption[]{a) The I$^*$ graph $n_1 =2$, $ n_2 = n_3 = n_4 = n_5 = 1$ drawn on the disk with handle.  b) The same graph drawn in the usual double line notation.  c) The same graph drawn as a torus with a hole.} 
\end{figure}

The integers $n_i$ are subject to several parity constraints.  Because the operators $A$ and $A^\dagger$ alternate along the fundamental line, there must be an even number of propagators attaching between the ends of any given propagator.  By applying this to given propagators of types 1, and 2, we find that $n_2 + n_3$ and $n_1 + n_3 + n_4$ must be even.  Type 3 propagators give the sum of these conditions, type 4 give the same condition as type 1, and type 5 give no conditions.  This leaves 
four cases:
\begin{equation}
(n_1, n_2, n_3, n_4) = OOOE, \ EOOO,\ OEEO,\ EEEE\ , \label{parity}
\end{equation}
($E = $ even, $O=$ odd), and in each case $n_5 = 0$ or 1.

It remains to determine the number of backward propagators $b$, since each of these brings in a factor of $y$.  The number is approximately half the total number of propagators, but a precise count requires us to examine separately each parity case (\ref{parity}).  For $n_5 = 0$ one finds respectively
\begin{equation}
b = (n+1)/2\ ,\quad (n+1)/2\ ,\quad (n+2)/2\ ,\quad n/2\ ,
\end{equation}
where $n= n_1 + n_2 + n_3 + n_4$.  For $n_5 = 1$,
\begin{equation}
b = (n+1)/2\ ,\quad (n+1)/2\ ,\quad n/2\ ,\quad (n+2)/2\ .
\end{equation}

Recall that the weight~(\ref{weight}) of a given graph is $w^v y^b$, with $v = n + n_5$. 
We sum over the integers $(n_1,n_2, n_3,n_4)$ with the limits 
$n_1 \ge 1$, $n_2 \ge 1$, $n_3 \ge 0$, $n_4 \ge 0$, and also over $n_5$, separating according to parity case:
\begin{eqnarray}
\label{finalGone}
N^2 [S_{\rm I^*} \tilde G ]_{g=1} 
&=&  \sum_{OOOE} (w^n + w^{n+1})  y^{{(n+1)/2}} 
 +  \sum_{EOOO}  (w^n + w^{n+1}) y^{{(n+1)/2}}   \nonumber \\
&&  + \sum_{OEEO} (w^n y^{{(n+2)/2}} + w^{n+1} y^{{n/2}})
+  \sum_{EEEE }  ( w^n y^{n/2} + w^{n+1}  y^{{(n + 2)/2}} )
\nonumber \\
 &=& \frac{ w^3 y^2 (1 + w)^2 (1 + wy)}{(1-w^2 y)^4}  \nonumber\\
 &=& \frac{ x^3 y^2 (1-x)^2(1 - x [1-y ]) } {(1 - 2 x + x^2 [1-y])^4}
 \ . 
\end{eqnarray}
Here $x = w/(1+w) = -i \lambda \tilde{G}(T, \omega) /(1-y) \equiv  -i \lambda _y\tilde{G}(T, \omega) $.

Expand the correlator in $1/N^2$,
\begin{eqnarray}
\tilde{G}(T, \omega) = \tilde{G}^{(0)} (T, \omega) + { 1 \over N^2} \tilde{G}^{(1)} (T, \omega)+ {\cal{O}}\left({1 \over N^4}\right) \,, 
\end{eqnarray}
where $\tilde{G}^{(0)} (T, \omega)$ is identified with (\ref{Gzerosol}).  
Also we define $x_0 = -i \lambda _y\tilde{G}^{(0)}(\omega) $. 
The $1/N^2$ term in the SD equation is then
\ba
\tilde G^{(1)}= \f{i}\omega \left( -i \lambda_y \tilde G^{(1)} 
\partial_x [S_{\rm I^*} \tilde G ]_{g=0}
+ [S_{\rm I^*} \tilde G ]_{g=1}
\right)\Bigr|_{x = x_0} \ ,
\ea
where $[S_{\rm I^*} \tilde G ]_{g=0} = wy = xy/(1-x)$ from the single planar I* graph.
The solution is 
\begin{equation} 
\label{finalsol}
\tilde{G}^{(1)}(T,\omega) =
\frac {i y^2 x_0^3  (1 -  x_0)^4 (1 - x_0 [1-y ] ) }
 {  (1 - 2 x_0 + x_0^2 [1-y])^4   (\omega [1-x_0]^2 - \lambda_y y)}\ .
 \end{equation}

The RHS of Eq.~(\ref{finalsol}) is a rational function of $\omega$ and of $x_0$, so its branch cut comes only from that of $x_0$: it is in the same place as for the planar amplitude $\tilde{G}^{(0)} (T, \omega)$.  For real $\omega$,  Re$\,\tilde{G}^{(1)}(T, \omega)$ is non-zero only if
Re$\,\tilde{G}^{(0)}(T, \omega)$ is also non-zero.  Furthermore the continuous spectrum of Re$\,\tilde{G}^{(0)}(T, \omega)$ is 
not modified by the leading perturbative $1/N^2$ correction Re$\,\tilde{G}^{(1)}(T, \omega)$. 
Note that from (\ref{genericSD}), (\ref{genericform}), each I$^*$ graph contributing to $S_{I^*} \tilde G$ 
has the same branch points as the planar amplitude, and the same continuous spectrum.  This need not hold after summing an infinite series of I $^*$ graphs, but we have found that $[S_{\rm I^*} \tilde G]_{g=1}$ (\ref{finalGone}) is a rational function of $x_0$ and $\omega$, and so does not introduce new cuts.  We expect that this will continue to hold at higher genus, so the
continuity of the spectrum and the positions of branch points will be the same at any finite order in $1/N^2$.

However, even this first correction does change the nature of the branch point.  By inserting the result~(\ref{Gzerosol}) for the planar propagator, one finds that $1 - 2x_0 + x_0^2 [1 - y]$ vanishes as $(\omega - \omega_\pm)^{1/2}$ at the ends of the cut, so that the denominator vanishes as $(\omega - \omega_\pm)^{2}$.  The branch point behavior of $\tilde{G}^{(1)}$ is then more singular than that of $\tilde{G}^{(0)}$: there is an $(\omega - \omega_\pm)^{2}$ double pole, as well as a subleading $(\omega - \omega_\pm)^{-3/2}$ branch cut, as opposed to $(\omega - \omega_\pm)^{1/2}$ in the planar term.  (This effect, which seems rather accidental in the present approach, will be more evident in the next section.)  This implies that  $G^{(1)}/G^{(0)}$ grows at long times.  In the conclusions we will discuss the possible relevance of this for the information problem.

It would be very useful to extend the graphical solution to all orders in $1/N^2$.  The main challenge seems to be the treatment of the modular group.  It would be interesting to find in particular some string field description, in which the sums over adjoint propagators can be thought of as a string propagator.
We hope to return to this in future work.


\sect{Loop approach}

One of our goals is to find a `bulk' description of our model, one analogous to the gravity side of gauge/gravity duality.  Bulk quantities are invariant under the gauge symmetries of the CFT, so we must work with $U(N)$ invariant objects.  Recasting gauge theory in terms of invariants has long been a tantalizing idea for connecting gauge theory with string theory~\cite{Polyakov:1979gp,Migdal:1984gj}, but it has been difficult to implement.  In our model, it turns out to be a useful way to calculate.

The invariants that we work with are ${\rm Tr} \,Q^{n}$, and generating functions for these.  We have to be careful about ordering because $Q_{ij} = A^\dagger_{ik} A_{kj}$ is a matrix whose elements are operators.  We specify the natural ordering whenever there is a matrix multiplication, e.g.
\begin{equation}
Q^n_{ij} = Q_{ik} Q_{kl} \ldots Q_{mj}\ .
\end{equation}
That is, we think of $Q$ as acting on the tensor product of the index space and the adjoint Hilbert space.
We will encounter expressions where other orderings arise, and for these we write the indices explicitly.  Also we define $\tilde Q_{ij} =  A_{ik} A^\dagger_{kj}$.  Note that the matrix elements $Q_{ij}$ commute with the $\tilde Q_{kl}$.

Because the thermal ensemble has
no fundamental excitation, we can write the correlator (\ref{correlator}) as
\ba
N G(t)=
\theta(t)\left\langle a_i 
e^{-i h a^\dagger _j Q_{jk} a_k t} a^\dagger_i\right\rangle_T.
\label{correlator-int}
\ea
(We will write $\langle \ldots\rangle_T$ simply as
$\langle \ldots\rangle$ hereafter.)
Since we consider the limit that the mass of fundamental field is very large, the 
number of fundamental field is always one.   
If we expand the exponential,
only adjacent oscillators can be contracted.
Thus we can express the correlator in terms of invariants as
\begin{eqnarray}
N G(t) &=& \theta(t) \left\langle {\rm Tr} \,e^{- i h Q t} \right\rangle\ , \nonumber\\
N \tilde G(\omega) &=& \left\langle {\rm Tr}\, \frac{i}{\omega - hQ}
\right\rangle\ .
\label{G-inv}
\end{eqnarray}
The trace is over the $N$-dimensional matrix indices, while the (thermal) expectation value is in the free adjoint Hilbert space. 

\subsection{Thermal loop equations}

In a thermal ensemble, the Heisenberg operators satisfy
\begin{equation}
\left\langle A_{ij}(-i\beta) {\cal O}_{ji}(0) \right\rangle = \left\langle  {\cal O}_{ji}(0) A_{ij}(0) \right\rangle \ ,
\end{equation}
for any ${\cal O}_{ji}$.
The free field equation $A_{ij}(t) = e^{-imt} A_{ij}(0)$ then implies
\begin{equation}
 \left\langle  {\cal O}_{ji} A_{ij} \right\rangle = y \left\langle A_{ij} {\cal O}_{ji} \right\rangle\ ,
 \label{oaao}
\end{equation}
or
\begin{equation}
 \left\langle  {\cal O}_{ji} A_{ij} \right\rangle =
\frac{y}{1-y}
\left\langle [ A_{ij} ,  {\cal O}_{ji} ]\right\rangle \label{loopcomm}
\end{equation}
 We now omit the common time argument 0 from all operators.  The `loop equation'~(\ref{loopcomm}) determines all expectation values iteratively, since the right side has two fewer mode operators than the left.

Let us first illustrate this with some simple examples.
For ${\cal O}_{ji} = A^\dagger_{ji}$ we obtain from Eq.~(\ref{loopcomm})
\begin{equation}
\left\langle {\rm Tr} \,Q \right\rangle = \frac{N^2 y}{1-y}\ .
\end{equation}
For ${\cal O}_{ji} = (A^\dagger A A^\dagger)_{ji}= (Q A^\dagger)_{ji}$,
\begin{equation}
\left\langle {\rm Tr} \,Q^2 \right\rangle = \frac{Ny}{1-y} \left\langle {\rm Tr} \,(\tilde Q + Q) \right\rangle = \frac{Ny}{1-y} \left\langle {\rm Tr} \,(2Q )+ N^2 \right\rangle
= \frac{N^3(y + y^2)}{(1-y)^2}
\ .
\end{equation}
For ${\cal O}_{ji} = Q_{kk} A^\dagger_{ji}$,
\begin{equation}
\left\langle {\rm Tr} \,Q\, {\rm Tr} \,Q \right\rangle = \frac{y}{1-y} \left\langle {\rm Tr} \,\tilde Q \right\rangle + \frac{N^2 y}{1-y} \left\langle {\rm Tr} \,Q \right\rangle =
 \frac{N^2 y}{(1-y)^2} +  \left\langle {\rm Tr} \,Q \right\rangle^2 \ .
\end{equation}
In particular we obtain the connected contribution, 
which is down by $O(N^2)$.

Now let ${\cal O}_{ji} = (Q^n A^\dagger)_{ji}$; 
this will require a bit more work.
First, the loop equation gives
\begin{equation}
\left\langle {\rm Tr} \,Q^{n+1} \right\rangle = \frac{y}{1-y} \sum_{p=0}^n
\left\langle {\rm Tr} \,Q^{p} \, {\rm Tr} \, \tilde Q^{n-p} \right\rangle  \label{qn1}
 \ .
\end{equation}
If $Q$ were a $c$-number matrix, the cyclic property of the trace would give ${\rm Tr} \, \tilde Q^{r} = {\rm Tr} \, Q^{r}$, but here we have commutators,\footnote{
Alternatively, one can compute $\langle \Tr\, Q^{n+1}\rangle$,
and hence the correlator (\ref{G-inv}),
by applying both (\ref{loopcomm}) and
\ba
\langle {\cal O}_{ji}  A^\dagger_{ij}\rangle
=\oo{1-y}\langle [ {\cal O}_{ji}, A^\dagger_{ij}]\rangle
\label{2nd-loop}
\ea
recursively
to eliminate the rightmost oscillator at each step.
After all oscillators are contracted,
the resulting terms are
in one-to-one correspondence with the Feynman graphs for $\tilde G$.
Equations (\ref{loopcomm}) and (\ref{2nd-loop})
respectively generate  backward and forward propagators,
and reproduce the Feynman rules (\ref{backfor}).
The ends of propagators in a Feynman graph
appear in the opposite order to how the corresponding oscillators
are inserted.
}  
\begin{equation}
{\rm Tr} \, \tilde Q^{r} = A_{ij} Q^{r-1}_{jk} A^\dagger_{ki}
= {\rm Tr} \, Q^{r} + \sum_{q=0}^{r-1} {\rm Tr} \, Q^{q}\,  {\rm Tr} \, \tilde Q^{r - q - 1}\ .
\label{recur}
\end{equation}
To solve this, we introduce the resolvents  
\begin{equation}
f(z) =  \frac{1}{Q - z }\ ,\quad 
\tilde f(z) =  \frac{1}{\tilde Q - z}\ ; \quad
F(z) = {\rm Tr} \,f(z)\ , \quad \tilde F(z) = {\rm Tr} \,\tilde f(z)\ .
\end{equation}
Multiplying Eq.~(\ref{recur}) by $-z^{-1-r}$ and summing $r$ from $0$ to $\infty$ gives
\begin{equation}
\tilde F(z) = F(z) - F(z) \tilde F(z)\ ,
\end{equation}
so that
\begin{equation}
\tilde F(z) = \frac{F(z)}{1+  F(z)}\ .
\end{equation}
To apply this to the recursion~(\ref{qn1}), multiply by $-z^{-2-n}$ and sum, giving
\begin{equation}
\left\langle {F(z) + N/z}  \right\rangle = - \frac{y}{1-y} \left\langle F(z) \tilde F(z)  \right\rangle
=  -\frac{y}{1-y}\left\langle \frac{F(z)^2}{1 + F(z)}  \right\rangle \ . \label{rofz}
\end{equation}
This can be used immediately to obtain the planar propagator, but we first derive further results that will be useful beyond planar order.

The next case is ${\cal O}_{ji} =   ({\rm Tr} \,Q^n)\, (Q^p A^\dagger)_{ji}$.  The loop equation is
\begin{equation}
\frac{1-y}{y}
\left\langle {\rm Tr} \,Q^{n}\, {\rm Tr} \,Q^{p+1} \right\rangle =  \sum_{q=0}^p
\left\langle{\rm Tr} \,Q^{n}\,{\rm Tr} \,Q^{q} \, {\rm Tr} \, \tilde Q^{p-q} \right\rangle
+  \left\langle \, [ A_{ij}, {\rm Tr} \,Q^{n}] (Q^p A^\dagger)_{ji} \right\rangle
 \ . \label{dtr}
\end{equation}
To simplify the last term we use
\begin{eqnarray}
[ A_{ij}, {\rm Tr} \,Q^{n}] &=& \sum_{q=0}^{n-1} \,Q^{q}_{kj} (A Q^{n-q-1})_{ik}
\nonumber\\
&= &  \sum_{q=0}^{n-1} \,Q^{q}_{kj} (\tilde Q^{n-q-1} A)_{ik}  \nonumber\\
&=&  \sum_{q=0}^{n-1} \,\tilde Q^{n-q-1}_{il} Q^{q}_{kj} A_{lk}  \ .
\label{commaqn}
\end{eqnarray}
We have used the fact  that $AQ = A A^\dagger A = \tilde Q A$, and that the matrix elements of $Q$ and $\tilde Q$ commute.  If we could reverse the order of $Q^q_{kj}$ and $A_{lk}$ this would simplify.  Thus we proceed
\begin{equation}
Q^{q}_{kj} A_{lk} = (A Q^{q})_{lj} - \sum_{r=0}^{q-1} ({\rm Tr} \,Q^{r}) (A Q^{q-r-1})_{lj}\ .
\end{equation}
Now multiply by $-z^{-1-q}$ and sum.  Then
\begin{equation}
f(z)_{kj} A_{lk} = (1 + F(z)) (A f(z))_{lj}\ .
\end{equation}
To use this, multiply Eq.~(\ref{commaqn}) by $-z^{-1-n}$ and sum on $n$,
\begin{eqnarray}
[ A_{ij}, F(z)] &=& -  \tilde f(z)_{il} f(z)_{kj} A_{lk}
\nonumber\\
&=& - \tilde f(z)_{il}( 1 + F(z)) (A f(z))_{lj} 
\nonumber\\
 &=& - (1 + F(z)) (\tilde f(z)^2 A)_{ij}  \ , \label{afga}
 \end{eqnarray}
so
\begin{equation}
[ A_{ij}, F(z)] (Q^p A^\dagger)_{ji}  = - (1 + F(z)) \Tr\,
(\tilde f(z)^2 \tilde Q^{p+1})  \ .
\end{equation}
Now multiply the relation~(\ref{dtr}) by $z^{-1- n} w^{-2-p}$ and sum from $(n,p) = (0,0)$ to obtain
\begin{equation}
\frac{1-y}{y}
\left\langle F(z) (F(w) + N/w) \right\rangle =  -
\left\langle F(z) F(w) \tilde F(w) \right\rangle
- \left\langle (1 + F(z))  {\rm Tr}\,(\tilde f^2(z) [\tilde f(w) + 1/w]) \right\rangle
 \ .
\end{equation}
We can simplify further by using partial fractions, 
\begin{equation}
\tilde f^2(z)  [\tilde f(w) + 1/w] = \partial_z\left(\frac{\tilde f(z) - \tilde f(w)}{z-w} + \frac{\tilde f(z)}{w}\right)\ .
\label{fff}
\end{equation}
Then
\begin{equation}
\frac{1-y}{y}
\left\langle F(z) (w F(w) + N) \right\rangle =  -
w\left\langle F(z) F(w) \tilde F(w) \right\rangle
- \left\langle (1 + F(z)) \partial_z\left(\frac{z\tilde F(z) - w\tilde F(w)}{z-w}\right) \right\rangle
 \ . \label{ffofz}
\end{equation}

Finally, we consider a very general case ${\cal O}_{ji} = F(z_1) F(z_2) \ldots F(z_l) ( f(w) A^\dagger)_{ji}$.
The commutator~(\ref{afga}) can be written
\begin{equation}
A_{ij} F(z) = ({\cal F}(z)A)_{ij}\ ,\quad {\cal F}(z)_{kl} = F(z)\delta_{kl} - (1+ F(z))\tilde f^2(z)_{kl}\ .
\end{equation}
Then $ \left\langle  {\cal O}_{ji} A_{ij} \right\rangle = y \left\langle A_{ij} {\cal O}_{ji} \right\rangle$ becomes
\begin{eqnarray}
\left\langle F(z_1) F(z_2) \ldots F(z_l) (w F(w) + N)\right\rangle
&=&
y\left\langle A_{ij} F(z_1) F(z_2) \ldots F(z_l) (  A^\dagger \tilde f(w))_{ji} \right\rangle
\nonumber\\
&=&
 y\left\langle{\rm Tr}\, {\cal F}(z_1) {\cal F}(z_2) \ldots {\cal F}(z_l) ( w \tilde f(w) + 1) \right\rangle\,.\ \ \label{general}
 \end{eqnarray}
This is our most general form for the loop equation.
The trace acts on the matrix indices of ${\cal F}$ and $\tilde f$.  Note that the matrix elements of $\tilde f(z_i)$ commute with $F(z_j)$ and so can all be brought to the right.

By partial fractions the RHS can again be written in terms of expectation vales of $F$, and their derivatives.  For example, in the case that $z_1 = z_2 = \ldots = z_l $,
\begin{eqnarray}
\left\langle F(z)^l  (w F(w) + N)\right\rangle &=& y \sum_{m=0}^l  {l \choose m} 
(-1)^m \left\langle F(z)^{l-m} (1+ F(z))^m
\, {\rm Tr}\, \tilde f(z)^{2m}  ( w \tilde f(w) + 1) \right\rangle \nonumber\\
&=& y \left\langle F(z)^l 
 ( w \tilde F(w) + N) \right\rangle + y
 \sum_{m=1}^l  {l \choose m} 
\frac{(-1)^m}{(2m - 1)!} \times \nonumber\\
&&
\left\langle F(z)^{l-m} (1+F(z))^m \partial_z^{2m-1} \left(\frac{z\tilde F(z) - w\tilde F(w)}{z-w} \right)
\right\rangle\ .\qquad
\end{eqnarray}
Here we used 
\begin{eqnarray}
\tilde f^{2m}(z)  [w \tilde f(w) + 1] = \frac{1}{(2m - 1)!} \partial^{2 m -1}_z\left(\frac{z \tilde f(z) -w \tilde f(w)}{z-w} \right)\ .
\end{eqnarray}

\subsection{\texorpdfstring{$1/N^2$}{1/N2} expansion}

We have not yet made any use of large $N$.  In the 't Hooft limit, $Q,z,w$ are of order $N$, $F$ is of order $1$, and $f$ is of order $N^{-1}$.  
We therefore rewrite the loop equation in terms of 
$\tilde \phi(u) =N \tilde f(Nu)$, 
$\Phi(u) = F(Nu)$ and $\tilde \Phi(u)=\tilde F(Nu)$.  
Note that
\begin{equation}
{\cal F}(Nu)_{kl} = \Phi(u)\delta_{kl} - \frac{1}{N^2} (1+ \Phi(u))\tilde\phi^2(u)_{kl}\ ,
\end{equation}
so the second term is nonplanar.  Therefore from Eq.~(\ref{general})
\begin{equation}
\left\langle \Phi(u_1) \Phi(u_2) \ldots \Phi(u_l) (v \Phi(v) + 1)\right\rangle
=
 y\left\langle\Phi(u_1) \Phi(u_2) \ldots \Phi(u_l) ( v \tilde \Phi(v) + 1) \right\rangle  + O(1/N^2) \ ,
 \end{equation}
or, using $\tilde\Phi = \Phi/(1+\Phi)$,
\begin{equation}
 \left\langle \Phi(u_1) \Phi(u_2) \ldots \Phi(u_l) \left\{ v\Phi(v) + 1 - y - yv \frac{ \Phi(v) }{1 + \Phi(v)} \right\}\right\rangle
\stackrel{\rm planar}{=} 0\ . \label{planloop}
\end{equation}
In the planar limit the expectation value factorizes, and so we have
\begin{equation}
\label{loopplanareq}
v\Phi_0^2(v) + (1 - y)(1+v) \Phi_0(v) + (1-y) 
=0\ ,
\end{equation}
where $\Phi_0(v) =  \left\langle \Phi(v) \right\rangle_{\rm planar}$.  
This reproduces the planar result~(\ref{SDquad}) for $\Phi_0(v) = i \lambda G(T, v\lambda) $.

We have assumed large-$N$ factorization, but we should be able to derive it from the loop equations, since we have argued that these are complete.  We give a somewhat formal derivation, as follows.  In Eq.~(\ref{planloop}) let the $u_i$ all be equal to $v$.  By forming a power series we can conclude for any analytic function $\tau(\Phi(v))$ that
\begin{equation}
 \left\langle \tau(\Phi(v)) \left\{ v\Phi(v) + 1 - y - yv \frac{ \Phi(v) }{1 + \Phi(v)} \right\}\right\rangle
\stackrel{\rm planar}{=} 0\ . \label{test}
\end{equation}
The thermal ensemble produces some probability distribution for $\Phi(v)$.\footnote{Since $\Phi(u)$ is the generating function of all Casimirs $\Tr\, Q^n$,
this is precisely the probability distribution of Young tableaux 
discussed in Sec. \ref{sect-tableaux}.
It is concentrated on a typical tableau determined there. 
The explicit formulas for $\Tr\, Q^n$ in terms of tableaux can be found in~\cite{Cordes:1994fc}. 
}
  Since Eq.~(\ref{test}) holds for arbitrary $\tau$, it must be that the distribution is concentrated on the zeros of the expression in the bracket.  There are two such zeros, but these have different large $v$ behaviors, $O(1/v)$ and $O(1)$.  Using the known asymptotic behavior we can conclude that the distribution is a delta function on the first solution.

We can now solve iteratively for the higher orders, expanding the general loop equation~(\ref{general}) as
\begin{eqnarray}
&& \left\langle \Phi(u_1) \Phi(u_2) \ldots \Phi(u_l) \left\{ v\Phi(v) + 1 - y - yv \frac{ \Phi(v) }{1 + \Phi(v)} \right\}\right\rangle
\nonumber\\
&& \qquad\quad=
- \frac{y}{N^2} \sum_{i=1}^l  \left\langle \Phi(u_1) \ldots (1 + \Phi(u_i)) \ldots \Phi(u_l) \,\partial_{u_i} \frac{ u_i \tilde \Phi(u_i) 
- v \tilde\Phi(v)}{u_i - v} \right\rangle
\nonumber\\
&& \qquad\qquad\,
+ \frac{y}{N^4} \sum_{1\leq i < j \leq l}  \left\langle \Phi(u_1) \ldots ( 1 +\Phi(u_i))\ldots (1 +\Phi(u_j)) \ldots \Phi(u_l) \vphantom{\frac{\Phi}{u}}\right. \times
\nonumber\\
&& \qquad\qquad\qquad \left.\times
\partial_{u_i} \partial_{u_j} \left\{ 
\frac{u_i \tilde \Phi(u_i)}{(u_i - u_j)(u_i - v)} + \frac{u_j\tilde \Phi(u_j)}{(u_j - u_i)(u_j - v)} + \frac{v\tilde \Phi(v)}{(v - u_i)(v - u_j)} \right\}  \right\rangle
\nonumber\\
 && \qquad\qquad\, + O\left( \frac{1}{N^6} \right)\ .
\end{eqnarray}
To obtain the propagator to $O(1/N^2)$ it suffices to take all $u_i$ to $v$ as in Eq.~(\ref{test}), giving
\begin{eqnarray}
&& \left\langle \tau(\Phi(v))
 \frac{[\Phi(v) - \Phi_0(v)]\,[v \Phi(v) - (1-y)/\Phi_0(v)]}{1 + \Phi(v)} 
\right\rangle
\nonumber\\
&&\qquad\qquad\qquad = - \frac{y}{2 N^2}  \left\langle \frac{d\tau}{d\Phi}(\Phi(v)) (1 + \Phi(v)) (2\tilde \Phi'(v) + v \tilde \Phi''(v)) \right\rangle \label{1n2}
\end{eqnarray}
The LHS is the same as in Eq.~(\ref{test}), but now written in terms of the planar solution
\begin{equation}
\Phi_0(v) = -\frac{(1-y)}{2v}\left( 1 + v - \sqrt{(v-v_+)
(v-v_-)} \right)\ ,\quad
v_{\pm} = \frac{ 1+ y \pm 2 \sqrt{y} }{1-y}\ ,  
\end{equation}
by using Eq.~(\ref{loopplanareq}). 
To solve this to order $1/N^2$, simply choose
\begin{equation}
\hat\tau(\Phi) =
\frac{1 + \Phi} {v \Phi - (1-y)/\Phi_0}\ .
\end{equation}
Inserting the planar solution $\Phi_0(v)$ on the RHS, Eq.~(\ref{1n2}) becomes
\begin{equation}
\Phi(v) = \Phi_0(v) - \frac{y}{2 N^2}  \frac{d\hat\tau}{d\Phi}(\Phi_0(v)) (1 + \Phi_0(v)) (2\tilde \Phi_0'(v) + v \tilde \Phi_0''(v)) + O\left( \frac{1}{N^4} \right) \,. \label{corr-loop}
\end{equation}
We can eliminate $\tilde \Phi_0$ in favor of $\Phi_0$
using $\tilde \Phi_0=\Phi_0/(1+\Phi_0)$.
Now derivatives of the quadratic equation (\ref{loopplanareq})
allow us to eliminate $\Phi_0'$ and $\Phi_0''$.
The $O(1/N^2)$ term in (\ref{corr-loop})
is then a rational function of $\Phi_0$, and its
equality with graphical approach result (\ref{finalsol}) can be shown by using the quadratic
equation again.


This representation of the solution makes evident the fact that the branch cut in the propagator becomes more singular: the second derivative converts the $(\omega - \omega_{\pm})^{1/2}$ to $(\omega - \omega_{\pm})^{-3/2}$.  This still misses the full singularity, as there is an additional $(\omega - \omega_{\pm})^{-1/2}$ hidden in $d\tau/d\Phi$.   It is also evident, from the form of the higher nonplanar terms, that there there will be two additional derivatives at each further order in $1/N^2$, and so the singularity will become two powers worse at each successive order.
Therefore at late times, the non-planar corrections $G^{(g)}(t)$ grow as $t^{2g-3/2}$.

\sect{Solution by summing over tableaux}
\label{sect-tableaux}

For  a space-time picture to emerge from a quantum mechanical model,
an important step would be to find variables that
are to be path-integrated and identified with the metric.
It is conceivable that such variables are
in fact discrete at the fundamental level, forming
 continuum approximately only in the large $N$ limit.
Here we use group theory and propose one such description for the $qQ$ model
(\ref{tilde}); the sum over Young tableaux
replaces the path-integral over the metric.
In the large $N$ limit, we recover the expression for $\tilde G$
obtained in Sec.~2 and 4. 


\subsection{Sum over tableaux}



We start with the Fourier transform of (\ref{correlator-int})
\ba
N \tilde G(\omega)=
\left\langle a_i \f{i}{\omega+h q\cdot Q} a^\dagger_i\right\rangle,
\ea
where $q\cdot Q=q_{ij}Q_{ji}$.
By inserting the completeness relation in the middle,
we can write the correlator as 
a sum
\ba
-i\tilde G(\omega)&=&
\oo N (1-y)^{N^2}\sum_B 
\langle B|
 \f{y^{\Tr\, Q}}{\omega+h q\cdot Q} |B \rangle.
\label{sum-B}
\ea
The sum is over the states $|B\rangle$
with one fundamental excitation ($a^\dagger a=1$).
Such states form a large representation of the $U(N)\times U(N)$ acting on the left and right indices of $A^\dagger$; we now decompose it into irreducible ones.

The states $|B\rangle$ are obtained by acting
one $a^\dagger_i$ and several $A^{\dagger}_{ij}$, so they 
form the representation
\ba
\Yboxdim{7pt}
(\yng(1)\hspace{1pt},1)\otimes\mathop{\oplus}_{l=0}^\infty {\rm Sym}^l({\bf \bar{\yng(1)}\hspace{1pt},
\yng(1)})
\ea
of $U(N)\times U(N)$.
We symmetrize because $A^\dagger$ are bosonic.  To decompose the symmetric product we compute the character, for $(U,V)\in U(N)\times U(N)$.  Denoting by $\vec k = (k_1, k_2, \ldots)$ an infinite series of non-negative integers, we have
\ba
\Yboxdim{5pt}\sum_l \Tr_{{\rm Sym}^{l}(\bar{\yng(1)},\yng(1))}(U^\mo,V)
&=& \sum_l \frac{1}{l!}
\langle 0|A_{ i_ln_l}\ldots A_{i_1n_1 }
U_{ n_1 m_1} \ldots  U_{ n_l m_l}
V_{j_1i_1}\ldots V_{ j_l i_l}
A^\dagger_{m_1 j_1}
\ldots 
A^\dagger_{m_l j_l}
|0\rangle
\nn\\
&=&
\sum_{\vec k} \prod_j \frac{1}{j^{k_j} k_j!} (\Tr\, U^j)^{k_j}
 (\Tr\, V^j)^{k_j}.
\label{character}
\ea
To rewrite this, recall that the vector $\vec k$ labels
a conjugacy class $C(\vec k)$
of the permutation group $S_l$ such that $l=\sum j k_j$.
Each irreducible representation of $S_l$ is
labeled by a Young tableau $R$ whose number of boxes
$|R|$ equals $l$.
The tableau also specifies an irreducible representation
of $U(N)$.
The Frobenius relation 
\ba
\prod_j (\Tr U^j)^{k_j}=\sum_R \chi_R(C(\vec k))\Tr_R U
\ea
and the orthogonality 
\ba
\sum_{\vec k}\prod_j \oo{j^{k_j}k_j!}
\chi_R(C(\vec k))\chi_{R'}(C(\vec k))=\delta_{RR'},
\ea
where $\chi_R(C(\vec k))$ is a character for $S_l$ and $\Tr_R\, U$
is the trace of $U$ in the representation of $U(N)$
specified by $R$,
allow us to rewrite (\ref{character}) as
\ba
\Yboxdim{5pt}\sum_l \Tr_{{\rm Sym}^{l}(\bar{\yng(1)},\yng(1))}(U^\mo,V)
&=&\sum_R \Tr_R\,  U\, \Tr_R\, V=\sum_R \Tr_{\bar R}\,  U^\mo\, \Tr_R\, V.
\ea
Thus
\ba
\Yboxdim{7pt}
\mathop{ \oplus}_{l=0}^\infty {\rm Sym}^l({\bf \bar{\yng(1)}}\, , {\bf 
\yng(1)})
=\mathop{\oplus}_R (\bar R, R). \label{sym-decomp}
\ea
The sum is over all Young tableaux with at most $N$ rows.
Let us recall
 that a Young tableau $R$ 
represents a partition 
 $l=R_1+R_2+\ldots+R_N$ of an integer $l\equiv|R|$
by a non-decreasing non-negative integers  $R_1\geq R_2\geq \ldots R_N\geq 0$;
is visualized by a collection of boxes where the $i$-th row
has $R_i$ boxes; and
specifies an irreducible representation of $U(N)$ as in (\ref{sym-decomp}).
The box in the $i$-th row and $j$-th column will be denoted by $(i,j)$.

We further decompose
$\Yboxdim{7pt}\yng(1)\otimes \bar R$ using
the Littlewood rule.
Let us consider 
a large rectangle that contains $R$ and has $N$ rows.
The complement of $R$ is another Young tableau.
The Littlewood rule instructs us to add a box to the complement so
that the result is again a Young tableau.
This is equivalent to removing a box $(i,R_i)$
at an outward corner of $R$, resulting in
a Young tableau $R'=R-(i,R_i)$ with $R'_j=R_j-\delta_{ji}$.
For $i=N$, we relax the conditions on $R'$ so that $R'_N$ can be $-1$.
Then
\ba
\Yboxdim{7pt}
\yng(1)\otimes \bar R=\mathop\oplus_{(i,R_i):{\rm corner}} \overline{ (R-(i,R_i))}.
\ea

Thus, the Hilbert space decomposes as
\ba
\Yboxdim{7pt}
(\yng(1)\hspace{1pt},1) 
 \otimes\mathop{\oplus}_R (\bar R, R)
=\mathop{\oplus}_R \mathop\oplus_{(i,R_i):{\rm corner}}
(\overline{ R-(i,R_i)},R).
\ea
In the first expression, the generators $q$ and $Q$ act only on 
$\Yboxdim{7pt}\yng(1)\hspace{1pt}$ and $\bar R$, respectively.
In the second, $q+Q$ act on
$\overline{ R-(i,R_i)}=\bar R'$.
Writing the interaction in terms of the quadratic Casimirs
\ba
-h q\cdot Q=\half h\Tr \left(q^2+Q^2-(q+Q)^2 \right)
\ea
and using their values~\cite{Cordes:1994fc, Fulton-Harris}
\ba
\Tr\, q^2|_{\Yboxdim{5pt} \yng(1)}=N,~~~~~
\Tr\, Q^2|_{\bar R}=\sum_{j=1}^N R_j(R_j-2j+1+N),~
\nn\\
\Tr(q+Q)^2|_{\bar R }=\sum_j R'_j(R'_j-2j+1+N),~~~~~
\ea
we find the energy spectrum $-h q\cdot Q=h(N-i+R_i).$
Noting $\Tr\, Q|_{\bar R}=|R|$, we can finally rewrite (\ref{sum-B}) as
\ba
-i \tilde G(\omega)&=&
(1-y)^{N^2}
\sum_{R} y^{|R|} (\dim R)^2
\ \Omega(\omega)=\langle \Omega(\omega)\rangle, \label{G-as-sum2}
\ea
where we have defined
\ba
\Omega(\omega)&\equiv&\sum_{(i,R_i):{\rm corner}}
\f{ \Delta_i }{\omega- h(N-i+R_i)},\\
\Delta_i&\equiv& \oo N \f{\dim (R-(i,R_i))}{\dim R},
~~\sum_{(i,R_i):\rm corner}\Delta_i=1.
\ea
In this form, the correlator is given as the average 
of the ``resolvent''
$\Omega(\omega)$
in the statistical ensemble of Young tableaux with weights
$\propto y^{|R|}(\dim R)^2$.

\subsection{Typical tableau at large \texorpdfstring{$N$}{N}}
Statistics of tableaux with various weights have been studied
in the mathematics literature. See~\cite{Okounkov} and references therein. 
In physics, similar problems appeared in studying the Seiberg-Witten
theory using instantons~\cite{Nekrasov:2003rj}.

\begin{figure}
\psfrag{(x, y)}{$(x,y)$}
\psfrag{(0, 0)}{$(0,0)$}
\psfrag{1}{$1$}
\psfrag{u}{$u$}
\psfrag{v}{$v$}
\psfrag{x}{$x$}
\psfrag{y}{$y$} 
\center \includegraphics[width=10cm]{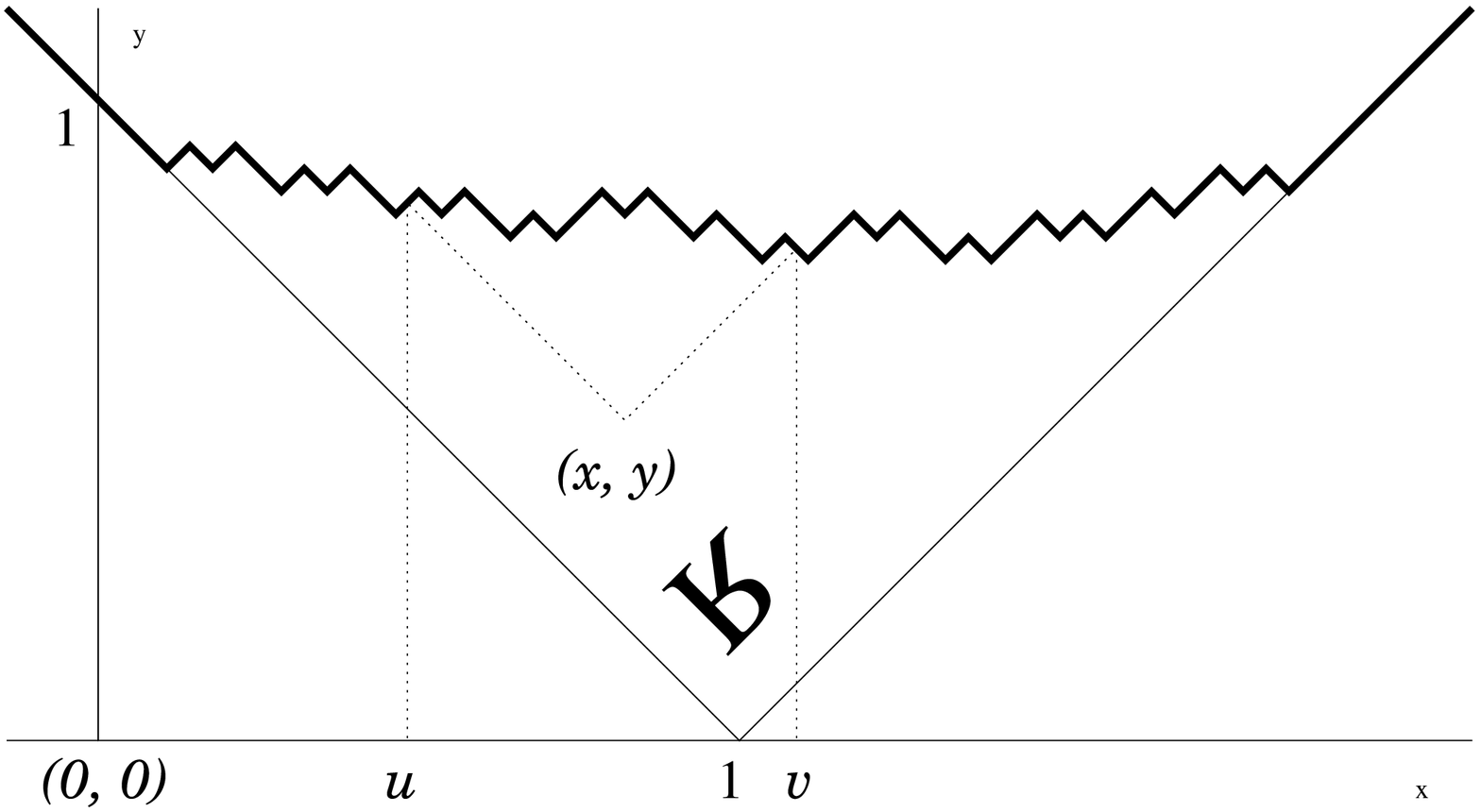}     
\caption[]{The Young diagram $R$, shown rotated and inverted, 
defines a piecewise linear
curve $y=f(x)$.
The upper corner of the box $(i,j)$ is placed at $(x,y)=(1+(j-i)/N, (j+i)/N)$.
The hook length is proportional to $v-u$.
} \label{young-f}
\end{figure}

As shown in Fig. \ref{young-f}, each tableau $R$ specifies a piecewise linear curve $y=f(x)$
such that $f(x)\geq |x-1|$.
Since $R$ can have at most $N$ rows, we must have $f(x)=-x+1$ for $x\leq0$.
The number of boxes $|R|$ is proportional to the
area of the region $|x-1|<y<f(x)$.
The dimension of the $U(N)$ representation is given by the formula
\cite{Fulton-Harris}
\ba
\dim R=\prod_{(i,j)\in R} \f{N+j-i}{h(i,j)},
\ea
where the hook length $h(i,j)$ for the box $(i,j)$ is
defined as
\ba
h(i,j)=R_i-j +R^T_j -i+1
\ea
with $R_i$ and $R^T_j$ being the number of
boxes in the $i$-th row and $j$-th column, respectively.
The box is mapped to  $(x,y)=(1+(j-i)/N, (j+i)/N)$,
so when $N$ is large, 
$\dim R$ is approximately given by
\ba
\dim R= \exp \f{N^2}2 \int_{|x-1|<y<f(x)} dx dy 
(\log x-\log(v-u)).
\ea
The quantities $u$ and $v$ are given by $u=x-f(u)+y$
and $v=x+f(v)-y$, and are indicated  in Fig.~\ref{young-f}. 
Thus the weight in (\ref{G-as-sum2}) is now given by
\ba
y^{|R|}(\dim R)^2
=e^{-N^2 S[f]},
\ea
where we have defined the ``action'' $S[f]$ by
\ba
2S[f]
&=& \int dx \left( \f{m}T-2\log x\right)\left(f(x)-|x-1|\right)
\nn\\
&&~~~~~+
\int_{u<v}d u d v (1+f'(u))(1-f'(v))\log (v-u).
\ea

In the large $N$ limit, the ensemble
is expected to be dominated by
the tableaux whose shapes
are almost identical.
In other words, the probability distribution
for $f$ should be sharply peaked
around some $f_*$, which specifies the shape of a typical tableau.
We assume that $f_*(x)$ is convex, so that
$-1<f'_*(x)<+1$ for $x_-<x<x_+$ and
$f'_*(x)=\pm 1$ for $x \gtrless x_\pm$.
The values of $x_\pm$ will be determined below.

To find $f_*$ as a 
minimum of $S[f]$, let us take the variation
\ba
\delta S[f]
&=&
-\half \int_{-\infty}^\infty dx\ \delta f'(x)\left(
\f{m}T x-2x\log x+2x +I[f](x)
\right)
\nn\\
&=&
\half \int_{-\infty}^\infty dx\ \delta f(x)\left(
\f{m}T -2\log x+I[f]'(x)
\right)
,
\ea
where we have defined the integral transform
\ba
I[f](x)\equiv \int_{-\infty}^\infty dx' \left( (x-x')\log |x-x'|-(x-x')\right)
f''(x').
\ea
We see that the function $f_*(x)$ must satisfy the following
conditions to be a minimum of $S[f]$:
\ba
\f{m}T-2\log x
+I[f_*]'(x)
\left\{
\begin{array}{lll}
\geq 0 &\hbox{ when } &x >x_+ \hbox{ or } x<x_-,
\\
=0 &\hbox{ when} & x_-\leq x\leq x_+.
\end{array}
\right.\label{SPE}
\ea
For $f_*$ to be a minimum,
$\delta S$ has to be zero or positive
under the variation $\delta f$ which is constrained
as $\delta f(x) \geq 0$
when $f'(x)=\pm 1$, leading to the inequality.

To find such $f_*(x)$, consider an analytic function $\varphi(z)$
defined on the upper half-plane as
\ba
\varphi(z)\equiv
1+\f{i}\pi\left[ \f{m}T-2\log z+ \int_{-\infty}^{\infty}
 dx
\log(z-x)
 f''_*(x)\right].
\ea
Since $\log(x+i\epsilon)=\log|x|+\pi i \theta(-x)$,
the restriction to the real axis is
\ba
\varphi(x)\equiv \lim_{\epsilon \ra 0}\varphi(x+i\epsilon)=
f'_*(x)+2\theta(-x)+
\f{i}\pi \left(
I[f_*]'(x)+\f{m}T-2\log |x|\right).
\ea
Then we see from (\ref{SPE}) that the conformal map $\varphi$
takes the interval $[x_-, x_+]$
to $[-1,1]$.
The interval  $[0, x_-]$
is mapped to the vertical line passing the point $z=-1$,
while the half lines $[x_+, +\infty)$
and $(-\infty, 0]$
are
are mapped to
the vertical line passing $z=1$.
Because of the logarithm we have
$\displaystyle\lim_{\epsilon\ra 0}\varphi(\pm \epsilon)=
\pm 1 +i \infty$, and the large $|x|$ behavior
\ba
I[f_*]'(x)=2\log |x|-\f{2}x+\left(1+\f{2|R|}{N^2}\right)\oo{x^2}+{\cal O}\left(
\oo{x^3}\right),\label{I-asymp}
\ea
implies that 
\ba
\varphi(z)=1+i\f{m}{\pi T}-\f{2i}\pi \oo z+{\cal O}
\left(\oo{z^2}\right) \hbox{ as } z\ra \infty.
\label{phi-exp}
\ea
In particular, $\varphi(\infty)=1+i m/\pi T$.

\begin{figure}
\begin{center}
\begin{tabular}{ccc}
\psfrag{z=0}{$0$}
\psfrag{x-}{$x_-$}
\psfrag{x+}{$x_+$}
 \includegraphics[scale=.4]{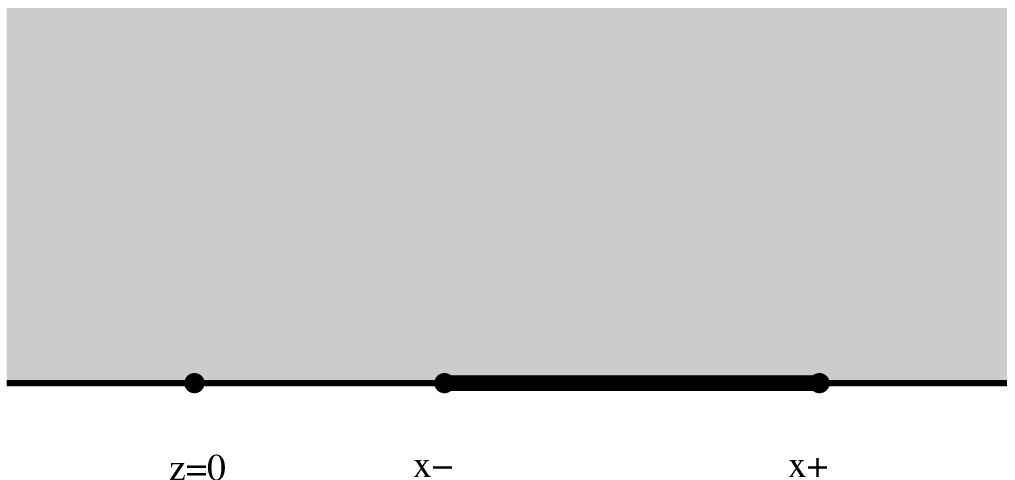} 
~~~
&
\,
\psfrag{+1}{$+1$}
\psfrag{-1}{$-1$}
\psfrag{a}{$
\varphi_1(\infty)$}
\includegraphics[scale=.4]{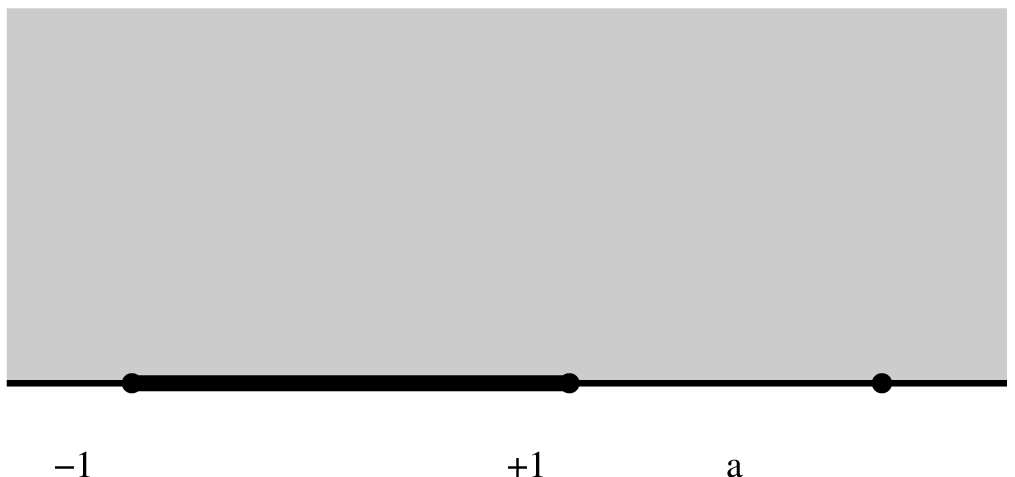} 
~~~
&
~~~
\psfrag{+1}{$+1$}
\psfrag{-1}{$-1$}
\psfrag{point}{$\displaystyle{1+i\f{m}{\pi T}}$}  
\includegraphics[scale=.4]{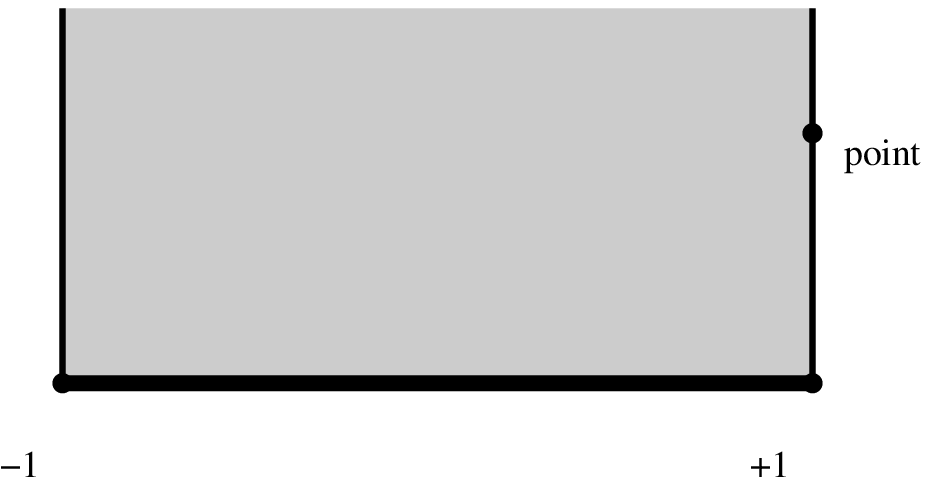} 
\\ 
a)&b)&c)
\end{tabular}
\end{center}
\caption[]{a)
The upper half $z$-plane on which the function
$\varphi(z)$ is defined.
b) An auxiliary half-plane we introduce to determine $\varphi(z)$.
It is useful to map the $z$-plane
to this half $w$-plane by a M\"obius transformation, bringing
$x_\pm$ to canonical locations $\pm 1$.
c) The image of the half $z$-plane by $\varphi$.
} \label{conformal-map} 
\end{figure}

These conditions can be visualized by saying
that $\varphi$ maps the
upper half-plane in Fig. \ref{conformal-map}a)
to the semi-infinite strip in Fig. \ref{conformal-map}c).
The four points $\{0, x_\pm, \infty\}$
should be mapped to $\{\infty, \pm 1, 1+im/\pi T\}$ respectively.
To determine $\varphi$, it is useful
to first consider 
a M\"obius transformation
that maps the three points $0$ and $x_\pm$ to $\infty$ and $\pm 1$
(see Fig. \ref{conformal-map}b))
\ba
\varphi_1(z)=
\f{(x_+ + x_-)z - 2 x_+ x_-}{(x_+-x_-)z}.
\ea
The conformal transformation from the half-plane
to the strip that respectively maps   $\infty$ and $\pm 1$
to $+i\infty$ and $\pm 1$
is uniquely given by $\varphi_2(w)=(2/\pi)\arcsin(w)$.
Therefore we conclude that
$\varphi$ is given by the composition $\varphi=\varphi_2\circ \varphi_1$.

We still need to determine $x_\pm$.
The map of the fourth point ($\infty$) imposes one condition on $x_\pm$:
\ba
\sin\left[\f{\pi}2\left(1+i\f{m}{\pi T}\right)\right]
=\varphi_1(\infty)=\f{x_+ + x_-}{x_+ - x_-}.
\ea
We get the other condition 
by comparing the expansion
of $\varphi_2 \circ \varphi_1$ around $z=\infty$
with (\ref{phi-exp}).
Equating the $O(1/z)$ terms, we find that
\ba
x_+ x_-=1.
\ea
Thus $x_\pm$ are given by
\ba
x_+=\coth \f{m}{4T},~~x_-=\tanh \f{m}{4T},
\ea
and the conformal map $\varphi$ by
\ba
\varphi(z)=\f{2}\pi \arcsin\left(
\cosh\f{m}{2T}-\oo z \sinh\f{m}{2T}
\right).
\ea


Finally, the dominant tableau has the shape given by
\ba
f'_*(x)
=\f{2}\pi \arcsin\left(
\cosh\f{m}{2T}-\oo x \sinh\f{m}{2T}
\right)
\ea
within the interval $[x_-, x_+]$, and $f(x) = |x-1|$ outside. 
If one uses this to compute the $O(1/x^2)$ term in $I[f_*]'(x)$
and compare it with (\ref{I-asymp}),
one  finds that $|R|/N^2=y/(1-y)$ as expected from Bose statistics.

It is also possible to solve the conditions~(\ref{SPE}) by inversion of the finite Hilbert transform~\cite{Lechtenfeld:1992bb}.

\subsection{The planar correlator from the typical tableau}
\begin{figure}
\psfrag{y1}{$Y_1$}
\psfrag{y2}{$Y_2$}
\psfrag{y3}{$Y_3$}
\psfrag{y4}{$Y_4$}
\psfrag{y5}{$Y_5$}
\psfrag{x1}{$X_1$}
\psfrag{x2}{$X_2$}
\psfrag{x3}{$X_3$}
\psfrag{x4}{$X_4$}
\psfrag{x5}{$X_5\equiv 0$}
\center \includegraphics[width=10cm]{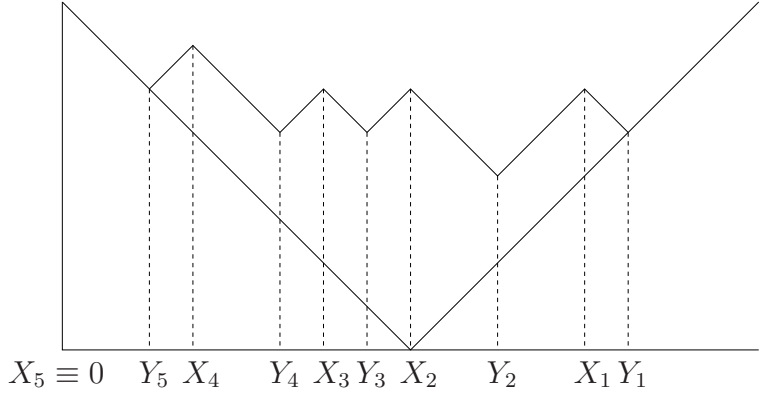} 
\caption[]{Symbols $X_I$ and $Y_I$ denote the values of $N-j+R_j$
at the $I$-th outward and inward corners.
} \label{corners} 
\end{figure}

To deal with the correlator (\ref{G-as-sum2}),  
we use the following trick~\cite{kerov-trans}.
Fix a tableau $R$ and suppose there are $a+1$ outward corners.
For our purpose there, the useful formula for the dimension is
\cite{Fulton-Harris}
\ba
\dim R=\prod_{1\leq i<j\leq N}
 \f{R_i-R_j+j-i}{j-i}.\label{dim-R}
\ea
If the box $(i,R_i)$ is at the $I$-the outward corner, the formula (\ref{dim-R})
gives  the residue of the pole
at $\omega=h(N-i+R_i)$ as
\ba
\Delta_i=\oo{N}\displaystyle\prod_{J=1}^{a+1}(Y_J-X_I)/\displaystyle\mathop{\prod_{J=1}^{a+1}}_{
J\neq I}^{}
(X_J-X_I). \label{Delta-i2}
\ea
Here $X_I$ and $Y_I$ denote the value of $N+R_j-j$ at the outward and inward corners as shown in Fig. \ref{corners}.
Note that $X_{a+1}\equiv 0$.
We recognize (\ref{Delta-i2}) as the coefficients in the partial fraction
expansion of 
\ba
\Omega(\omega)
=\f{1}{\lambda}-\oo{\lambda} \prod_{J=1}^{a+1} \f{h Y_J-\omega}{h X_J-\omega}.
\ea
The constant term is fixed by the asymptotics $\Omega(\omega)\sim 1/\omega$
for large $\omega$.
Because $f''(x)$ is a sum of delta functions at the corners,
the second term can be written as
\ba
&&\oo{ \omega}\exp\half \int_{0}^\infty dx f''(x)\log(x-\omega/\lambda)
\nn\\
&=&\f{\sqrt{(\lambda \alpha_+-\omega)(\lambda \alpha_- -\omega)}}{\lambda\omega}
\exp
\left(-\half \int_{\alpha_-}^{\alpha_+} dx f'(x)\oo{x-\omega/\lambda}\right).
\ea
Here we assumed that  $f'(x)=-1$ for $0<x<x_-$
and that  $f'(x)=1$ for $x>\alpha_+$.
Thus
\ba
\Omega(\omega)
=
\oo{\lambda}-\f{\sqrt{(\omega-\lambda \alpha_+)(\omega-\lambda
\alpha_-)}}{ \lambda \omega}
\exp
\left(-\half \int_{\alpha_-}^{\alpha_+} dx f'(x)\oo{x-\omega/\lambda}\right).
\label{G-from-R}
\ea
When $f(x)$ is a piecewise linear function representing $R$,
this relation is exact.

In the large $N$ limit, the sum over tableaux is dominated
by the saddle point $f_*$.
Thus one needs to perform the integral  in (\ref{G-from-R})
with $f=f_*$ and $\alpha_\pm=x_\pm$
to obtain the correlator.\footnote{
For this, change the variable to $\theta=(\pi/2)\varphi(x)$,
so that the integral takes the form $\int^{\pi/2}_{-\pi/2}
 d\theta\, \theta \cos\theta /(c_1-\sin\theta)(c_2-\sin\theta).$ 
It can be integrated by parts using $
\int^{\pi/2}_{-\pi/2} d\theta \log(c-\sin\theta)=\pi\log[(c+\sqrt{c^2-1})/2]
$.
}
One indeed finds that
(\ref{G-from-R}) agrees with the result (\ref{primed-G-SD})
obtained from the SD equation, namely,
\ba
\tilde G(\omega)=i\Omega(\omega)|_{f=f_*}=
 \frac{i(1-y)}{2\omega\lambda}\left( \lambda + \omega - \sqrt{(\omega-\omega_+)
(\omega-\omega_-)} \right).
\ea
Note that $\omega_\pm=\lambda x_{\pm}$.

\sect{Rectangular models}

The $qQ$ model can be generalized to the case that $A^\dagger_{ii'}$ is 
a rectangular $N \times K$ matrix, while the fundamental is still $N$-dimensional.  For example, in the limit of large $N$ with $K/N$ fixed, the graphs of Fig.~2 have weight $h^{n+1}KN^n$, and so the SD equation becomes
\begin{equation}
-i \omega \tilde G(T,\omega)= 1 - \frac{i y hK \tilde G(T,\omega)}{1 - y + i hN  \tilde G(T,\omega)}\ .
\end{equation}
The solution is now
\begin{eqnarray}
\tilde G(T,\omega) &=& \frac{i}{2\omega\lambda}\left[ h(N-Ky) + \omega (1-y) -(1-y) \sqrt{(\omega-\omega_+)
(\omega-\omega_-)} \right]\ , \nonumber\\
\omega_\pm  &=& \frac{h}{1-y}\left (N + yK \pm 2\sqrt{NKy}\right)
\ .  \label{neq1}
\end{eqnarray}

The sum over Young tableaux generalizes as follows,
\ba
-i \tilde G(\omega)&=&
(1-y)^{NK}
\sum_{R} y^{|R|} \dim_N R \dim_K R
\ \Omega(\omega)=\langle \Omega(\omega)\rangle, 
\\\Omega(\omega)&\equiv&\sum_{(i,R_i):{\rm corner}}
\f{ \Delta_i }{\omega- h(N-i+R_i)},\\
\Delta_i&\equiv& \oo N \f{\dim_N (R-(i,R_i))}{\dim_N R},
~~\sum_{(i,R_i):\rm corner}\Delta_i=1.
\ea
The subscripts $N,K$ refer to the dimensions of the representations of $U(N)$ and $U(K)$ with the given tableaux.  The tableaux thus have at most $\min(N,K)$ rows.

There are several special cases and limits that can be solved more fully.

\subsection{\texorpdfstring{$N = 1$}{N=1}}
Here,
\begin{equation}
H =  h a^\dagger a A^\dagger_{i'} A_{i'}
\end{equation}
is just the product of number operators.  The energy is 0 in the zero-fundamental sector 
and $h k$ in the one-fundamental sector, where $k$ is the number of adjoint excitations.  Thus,
\begin{eqnarray}
\tilde G(T,\omega) &=& \sum_{k=0}^\infty P(k) \frac{i}{\omega - h k}\ ,
\end{eqnarray}
where 
\begin{equation}
P(k) =  (1-y)^K y^k {k+K-1 \choose  k}
\end{equation}
is the normalized probability to find a total of $k$ excitations on $K$ oscillators.   Also,
\begin{equation}
G(t) =  \theta(t) \left( \frac{1 - y}{1-ye^{-i h t}} \right)^K\ . \label{nis1}
\end{equation}
Young tableaux are limited to a single row, with any number $k$ of boxes, so that $\dim_N R = 1$ and $\dim_K R =  {k+K-1 \choose  k}$, which reproduces the above result.

Though the exact solution
(\ref{nis1}) is available, it is interesting to consider a saddle
point approximation.
The dominant tableau at large $K$ is at
\begin{equation}
k_*/(k_*+K) = y\ ,\quad k_* = Ky/(1-y)\ .
\end{equation}
This simply moves the pole of the free propagator from $\omega = 0$ to 
$\omega = h K y/(1-y)$.  
To see the continuous spectrum we must look at the fluctuations around the dominant tableau,
\begin{equation}
(1-y)^K 
y^k {k+K-1 \choose  k} \approx 
\frac{e^{- (k-k_*)^2 /2K \gamma^2} }{\sqrt{2\pi K} \, \gamma}\ , \quad \gamma = \frac{\sqrt{y}}{1-y}\ .
\end{equation}
This has a width of order $\sqrt{K}$ so we can replace the sum by an integral, $k - k_* \to x\sqrt{K}$,
\begin{equation}
\tilde G(\omega) \approx i \int_{-\infty}^\infty \frac{dx}{\sqrt{2\pi}\gamma} \,
\frac{e^{-x^2/2 \gamma ^2}}{\omega - hK y/(1-y) - h  \sqrt K x + i\epsilon} \ .
   \label{wgauss}
\end{equation}
Since this has no singularities in the lower half-plane
except the pole,
$G(t)$ for $t>0$ can be obtained by a contour integral in $\omega$.
Integrating over $x$, one finds that the falloff in $t$ is gaussian.  
However, in the exact expression there are periodic recurrences at $\Delta t = 2\pi / h$.

We can also look at this graphically, as a large-$K$ vector model.  One would normally hold $h K$ fixed, and the dominant graphs are then tadpoles like the $n = 0$ term in Fig.~2.  In our model this is just an uninteresting shift of the frequency, as we see from the result~(\ref{wgauss}).  It is therefore more interesting to absorb this into an additive redefinition of $\omega$ and to take a different limit in which $\zeta = h \sqrt K$ is fixed.  The relevant graphs are like the $n=1$ term in Fig.~2, where the $K$-vector loop interacts at two points with the fundamental excitation.  For interaction times $t'$ and $t''$ this loop contributes
\begin{equation}
-\theta(t' - t'') \zeta^2 \gamma^2 \ , \label{2loop}
\end{equation}
so its integrated contribution is $- t^2\zeta^2 \gamma^2/2$.  For $N = 1$ 
the relative time-ordering of the interactions of different loops does not matter, and so the loops exponentiate,
\begin{equation}
G(t) = \theta(t) e^{ - t^2\zeta^2 \gamma^2/2 }\ .
\end{equation}
This is indeed the Fourier transform of the result~(\ref{wgauss}).  The falloff is gaussian in the large-$K$ approximation, while the recurrence time $\Delta t = 2\pi \sqrt{K} / \zeta$ grows as $\sqrt{K}$.

Of course, given the exact solution for $N=1$ we can write the full $1/K$ expansion, and it it useful to do so.  In particular for $t>0$
\begin{eqnarray}
\ln G(t ) &=& K \ln(1-y) - K \ln(1 - y e^{i \zeta t/\sqrt{K}})  \nonumber\\
&=&  K \ln(1-y) + K \sum_{r=1}^\infty \frac{y^r}{r} e^{i \zeta  t r/\sqrt{K}} \nonumber\\
&=& \sum_{r=1}^\infty \sum_{s=1}^\infty \frac{y^r}{r} \frac{ (i \zeta  t r)^s }{s!} K^{1 - s/2}\ .
\end{eqnarray}
For large order $s$, the dominant terms in the sum are at $r \sim s/|\ln y| $.
 Replacing the $r$ sum by an integral, the coefficient of $K^{1 - s/2}$ is approximately
\begin{equation}
\frac{1}{s^{3/2}} (i \zeta t / |\ln y|)^s\ .  \label{kexp}
\end{equation}
Thus, the $1/K$ expansion has a finite radius of convergence, but the radius $(m\beta/ \zeta t)^2$ decreases with $t$.  This is interesting in connection with the recurrences, as we will discuss further in the conclusions.

\subsection{Large \texorpdfstring{$K$}{K}, fixed \texorpdfstring{$N$}{N}}

We can extend the graphical solution to give the large-$K$ limit for any fixed $N$.  With $\zeta = h \sqrt K$ fixed, and the tadpole shifted away, the only graphs that survive as $K \to \infty$ are bubbles with two vertices, though at finite $N$ different bubbles can overlap in a nonplanar way.  
Thus we can replace $Q_{ij} =  A_{ik'}^\dagger  A^\vp_{k'j} $ with the $U(K)$ singlet field ${\varphi}_{ij} \sqrt{K}$, with two-point function
\begin{equation}
\left\langle {\rm T}\, {\varphi}_{ij}(t) {\varphi}_{kl}(t') \right\rangle \equiv \frac{1}{K}\left\langle {\rm T}\,{Q}_{ij}(t) {Q}_{kl}(t') \right\rangle
= \frac{y}{(1-y)^2} \delta_{il}\delta_{jk}\ .
\end{equation}
This is independent of time, so ${\varphi}_{ij}$ is governed by a gaussian matrix (rather than path) integral.  At fixed ${\varphi}_{ij}$ the fundamental propagator is $\frac{1}{N} {\rm Tr} (e^{-i \zeta \varphi t})$, and so
\begin{equation}
G(t) = \theta(t) I(t)/I(0),\quad I(t) = \frac{1}{N} \int d^{N^2} {\varphi}\, e^{-{\rm Tr}({\varphi}^2) /2 \gamma^2} 
{\rm Tr}(e^{- i \zeta{\varphi}t})\ .
\end{equation}
For $N=1$ this is an ordinary integral and reproduces the earlier result.  For $N=2$ we have the eigenvalue integral
\begin{equation}
I(t) \propto \int_{-\infty}^\infty d^2x\, (x_1 - x_2)^2 e^{-(x_1^2 + x_2^2)/2\gamma^2} (e^{- i \zeta  x_1 t} + e^{- i \zeta x_2 t}) \ .
\end{equation}
For general $N$ we can express $G(t)$ using 
the method of orthogonal polynomials~\cite{Drukker:2000rr}.

\subsection{\texorpdfstring{$y \to 1$}{y -> 1} limit, 
all \texorpdfstring{$N, K$}{N, K}}   

In general, the adjoint propagators form cycles with $k$ vertices as in Fig.~2, though at finite $N$ these cycles can overlap, and the time ordering need not match the cyclic ordering.  We would like to rewrite the cycle in terms a dual graph, as in Fig.~9.
\begin{figure}
\center \includegraphics[width=14cm]{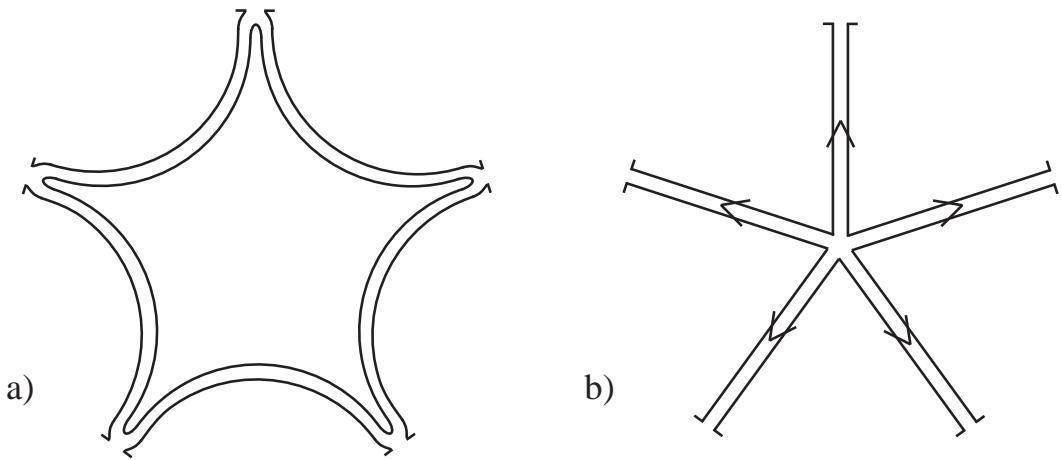}   
\caption[]{a) Loop of $k = 5$ vertices and adjoint propagators.  b) Dual graph with $\psi^{k}$ interaction.  The arrows point from $\psi$ to $\phi$.} 
\end{figure}
The loop becomes a vertex ${\rm Tr}\, \psi^{k}$, attached to the fundamental line by $\psi\varphi$ propagators.  We must add two fields, with an off-diagonal propagator, so that every propagator has one end on the central vertex and one on the fermion line.  

In general this seems rather complicated, because we must take account of the number of backward and forward propagators~(\ref{backfor}).  However, in the special case  $y \to 1$, where we must hold fixed $\chi = h/(1-y)$, these are equal and the total contribution of the vertices and adjoint propagators is simply $(-i\chi)^{k}K$.  
In this case, 
all graphs can then be summed via
\begin{eqnarray}
I(t) &=& \frac{1}{N} \int d^{N^2}\varphi \, d^{N^2}\psi \, e^{-{\rm Tr}(-i \varphi\psi + V(\psi) ) } \,
{\rm Tr}(e^{-i \chi \varphi t})\ , 
\nonumber\\
V(\psi) &=&  -K \sum_{k=1}^\infty \frac{ (-i \psi)^{k} }{k} = K \ln (1 + i \psi)\ .
\end{eqnarray}
The central vertex contributes $K (-i)^{k}$, the propagators contribute $i^k$, and the outer vertices contribute $(-i\chi)^k$.  The $U(N)$ group structure of the two graphs is also identical.

We can check the result for $N=1$, where it becomes an ordinary integral. The $\varphi$ integral produces $\delta(\psi - \chi t)$, and the $\psi$ integral then gives $e^{-V(\chi t)}$, which agrees with the $y \to 1$ limit of the earlier result~(\ref{nis1}).  Note that at infinite temperature we are accessing an infinite number of states, so there will not be recurrences.

\sect{Discussion}

The $qQ$ model, though in some ways quite simple, is seen to have a rich $1/N$ expansion.  Thus it may be a useful laboratory for studying this expansion in connection with AdS/CFT duality.  In particular, we have presented several ways of analysing the model  --- graphical, loop equations, Young tableaux --- each of which is suggestive of a bulk string or geometric language.

Our main interest in this model is that it captures the large-$N$ structure of the information problem, in that correlators decay to zero at infinite $N$ but not at finite $N$.  It is interesting from this point of view to consider the $N=1$ model of Sec.~6.1, which has the same features: there is gaussian decay in the large-$K$ limit but recurrences on a time scale $\sqrt{K}$.  In this case we have the whole $1/K$ expansion, and so we can analyze this further.  In fact, from the coefficient~(\ref{kexp}) we see that the expansion parameter is proportional to $t/\sqrt{K}$.  Thus there is no surprise that the limits $K \to \infty$ (giving decay) and $t \to\infty$ (giving recurrences) do not commute.

The existence of recurrences does not require this behavior of the perturbation theory.  Consider for example the function 
\begin{equation}
G_{\rm toy}(t) = \sum_{l = -\infty}^\infty e^{ - ( t - 2\pi l \sqrt{K}/\zeta)^2 \zeta^2 \gamma^2/2 }\ ,
\end{equation}
This has the same large-$K$ limit as the model, and recurrences on the same time scale, but by design there are no perturbative corrections at all.

The large-$K$ limit is a vector model, and there is no reason to expect the same result for matrix models.  However, it is notable that we have found that the expansion parameter in the matrix model is of order $t^2/N^2$.  We must be careful, however, in interpreting this.  Consider for example a system having a quasinormal mode $e^{-\Gamma t}$, with $\Gamma$ of order $T$.  Quantum gravity effects will certainly lead to a small correction to the decay rate, $\Gamma \to \Gamma + \Gamma_1/N^2$, so that the quasinormal behavior becomes
\begin{equation}
 e^{-\Gamma t}\left[ 1 - \Gamma_1 t /N^2 + O(t^2/N^4) \right]\ . \label{expdec}
\end{equation}
Thus, $1/N^2$ corrections that grow in time (relative to the leading term) are inevitable, even for a perfectly well-behaved perturbation expansion.\footnote{There is nothing particularly gravitational about this example: the same would hold for QED corrections to muon decay.}  Can we distinguish harmless growth from growth that leads to recurrences?  It may be a clue that the series~(\ref{kexp}) has a finite radius of converge, while the series~(\ref{expdec}) converges.  Thus it would be extremely interesting to determine at least the dominant large-$t$ behavior at each order in $1/N^2$, to see whether it represents a breakdown of the $1/N^2$ expansion at large times or simply a small shift in the leading behavior as above.  If there is a long-time breakdown of the $1/N^2$ expansion in this model, it is then important to determine whether it is an artifact of the model, or carries over to the full dual gauge theory.

In the context of real black holes, in order to understand possible corrections it is useful to consider Hamiltonian evolution along a series of spacelike surfaces (`nice slices') of small curvature (see, e.g., \cite{Lowe:1995ac}).  The absence of any large local curvature invariants then implies that gravity as a low energy effective theory does not signal its own break down due to high energy effects.  The dimensionless loop expansion factor is $G/R_{\rm s}^2$, where $R_{\rm s}$ is the Schwarzschild radius.  Dimensional analysis would still allow effects of order $G t /R_{\rm s}^3$, which become of order one on the lifetime of the black hole, and such effects were found in~\cite{Giddings:2007ie}.  As the above discussion shows, it is nontrivial to determine whether such large corrections signal a breakdown of the perturbative calculation, or something that is readily resummed.  In the calculation of Ref.~\cite{Giddings:2007ie} we believe it is the latter: the effect is an incremental change in the lifetime of a black hole due to the energy of a Hawking particle, similar to the incremental change~(\ref{expdec}).  Indeed, the no-hair theorem would seem to imply that secular effects can only manifest themselves through the mass (and other conserved charges) of the black hole.  Thus, from the Hawking radiation one can learn something about the amount of mass flowing into and out of the black hole over time, but one cannot distinguish two infalling objects of the same mass but different internal states.

\section*{Acknowledgments}

This work was supported in part by NSF grants PHY05-51164 and PHY04-56556.



\end{document}